\documentclass[10pt,letterpaper]{article}
\usepackage[top=0.85in,left=2.75in,footskip=0.75in]{geometry}

\usepackage{amsmath,amssymb}

\usepackage{changepage}

\usepackage[utf8x]{inputenc}

\usepackage{textcomp,marvosym}

\usepackage{cite}

\usepackage{nameref,hyperref}

\usepackage[right]{lineno}

\usepackage{microtype}
\DisableLigatures[f]{encoding = *, family = * }

\usepackage[table]{xcolor}

\usepackage{array}

\newcolumntype{+}{!{\vrule width 2pt}}

\newlength\savedwidth

\usepackage{setspace}

\raggedright
\usepackage[aboveskip=1pt,labelfont=bf,labelsep=period,justification=raggedright,singlelinecheck=off]{caption}

\makeatletter
\renewcommand{\@biblabel}[1]{\quad#1.}
\makeatother

\usepackage{lastpage,fancyhdr,graphicx}
\usepackage{epstopdf}
\fancyheadoffset[L]{2.25in}
\fancyfootoffset[L]{2.25in}
\newcommand*{\Method}{{\textsc{Skills Space}}\xspace}

\newcommand*{\PaperTitle}{{\textsc{Skill-driven Recommendations for Job Transition Pathways}}\xspace}

\usepackage[export]{adjustbox}
\usepackage{xspace}
\usepackage{subfig}
\usepackage{epstopdf}
\DeclareGraphicsExtensions{.png,.pdf,.eps,.jpg,.tif,.tiff,.ps}
\graphicspath{{img//}}

\usepackage[colorinlistoftodos,prependcaption,textsize=tiny,textwidth=10mm]{todonotes}  
\usepackage[nameinlink,capitalize]{cleveref} \usepackage{makecell}

\usepackage{multirow}
\usepackage{booktabs}
\usepackage{adjustbox}
\usepackage[section]{placeins}

\definecolor{navy}{rgb}{0.1, 0.1, 0.8}
\definecolor{gray}{rgb}{0.6, 0.6, 0.6}
\definecolor{myblue}{rgb}{.8, .8, 1}
\definecolor{olive}{rgb}{0.1, 0.5, 0.1}

\begin{document}
\vspace*{0.2in}

\begin{flushleft}

{\Large
\textbf\newline{Skill-driven Recommendations for Job Transition Pathways}
}
\newline
\\
Nikolas Dawson\textsuperscript{1,2*\textcurrency},
Mary-Anne Williams\textsuperscript{3},
Marian-Andrei Rizoiu\textsuperscript{4}
\\
\bigskip
\textbf{1} Centre of Artificial Intelligence, University of Technology Sydney, Sydney, Australia
\\
\textbf{2} OECD Future of Work Research Fellow
\\
\textbf{3} Business School, University of New South Wales, Sydney, Australia
\\
\textbf{4} Data Science Institute, University of Technology Sydney, Sydney, Australia
\\
\bigskip

\textcurrency Current Address: Centre of Artificial Intelligence, University of Technology Sydney, 15 Broadway, Ultimo NSW 2007, Australia

* nikolasjdawson@gmail.com

\end{flushleft}
\section*{Abstract}
Job security can never be taken for granted, especially in times of rapid, widespread and unexpected social and economic change.
These changes can force workers to transition to new jobs.
This may be because new technologies emerge or production is moved abroad. Perhaps it is a global crisis, such as COVID-19, which shutters industries and displaces labor \textit{en masse}. 
Regardless of the impetus, people are faced with the challenge of moving between jobs to find new work. Successful transitions typically occur when workers leverage their existing skills in the new occupation. 
Here, we propose a novel method to measure the similarity between occupations using their underlying skills. 
We then build a recommender system for identifying optimal transition pathways between occupations using job advertisements (ads) data and a longitudinal household survey. 
Our results show that not only can we accurately predict occupational transitions (Accuracy = 76\%), but we account for the asymmetric difficulties of moving between jobs (it is easier to move in one direction than the other). 
We also build an early warning indicator for new technology adoption (showcasing Artificial Intelligence), a major driver of rising job transitions.
By using real-time data, our systems can respond to labor demand shifts as they occur (such as those caused by COVID-19). 
They can be leveraged by policy-makers, educators, and job seekers who are forced to confront the often distressing challenges of finding new jobs.

\section*{Introduction}
In March 2020, COVID-19 caused entire industries to shutter as governments scrambled to `flatten the curve'. Jobs were lost or subject to an indefinite hiatus; firms went into `hibernation' to wait out the depressed demand; and governments exercised wartime measures of labor redeployment and wage subsidies of unprecedented scale. All in a matter of weeks.

Labor market shocks, such as those caused by COVID-19, force workers to abruptly transition between jobs.
Crises, however, are not the only cause of large-scale job transitions.
Structural shifts in labor demand are another major obstacle~\cite{acemoglu2011skills}, but usually unfold more gradually.
Indeed, technological advances were expected to cause the next wave of major labor market disruptions~\cite{brynjolfsson2014second, schwab2017fourth}. 
The `future of work' was to be defined by technologies like Artificial Intelligence (AI); technologies that would automate and augment workers, but at the same time transform the requirements of jobs and the demand for labor \textit{en masse}~\cite{Frey2017, acemoglu2018artificial, Frank2019-dn}.

Despite the impetus, many workers need to transition between jobs. 
In some countries, such as Australia, labor displacement has increased over the past two decades with relatively high levels of job transitions~\cite{sila2019job}; a situation exacerbated by COVID-19~\cite{Abs2020-ws}.
While job turnover is not innately negative and can be a signal of labor market dynamism, it does depend on how efficiently workers transition back into the workforce.
Transitioning from one job to another can be difficult or unfeasible when the skills gap is too large~\cite{nedelkoska2015skill}. Successful transitions typically involve workers leveraging their existing skills and acquiring new skills to meet the demands of the target occupation~\cite{poletaev2008human, gathmann2010general}. Therefore, transitioning workers successfully at scale requires maximizing the similarity between workers' current skills and their target jobs.
Skills, knowledge areas, and capabilities enable workers to achieve tasks required by jobs~\cite{nedelkoska2018}. 
We refer to these aspects of human capital as `skills' throughout this research, and we characterize labor market entities (individual jobs, standardized occupations, industries, etc.) as \textit{sets of skills}.

Here, we propose a novel method to measure the distance between sets of skills from more than 8 Million real-time job advertisements (ads) in Australia from 2012-2020. 
We call this data-driven methodology \Method. 
The \Method method enables us to measure the distance between any defined skill sets based on distances at the individual skill level.
When two skill sets are highly similar (for example, two occupations), the skills gap is narrow, and the barriers to transitioning from one to the other are low.
Drawing from previous work~\cite{poletaev2008human, gathmann2010general, nedelkoska2015skill, MovingBetweenJobs}, we construct a unique \emph{Job Transitions Recommender System} that incorporates the skill set distance measures together with other labor market data from job ads and employment statistics. 
This allows us to account for a wealth of labor market variables from multiple sources. 
The outputs of the recommender system accurately predict transitions between occupations (Accuracy = 76\%) and are validated against a dataset of occupational transitions from a longitudinal household survey~\cite{HILDA_2020}.
While previous studies have analyzed job transitions using the same or similar job ads data~\cite{Wef2018-eu, Wef2019-jv, Reskilling-AUS-2019, Kyle_Demaria2020-ec}, they have not accounted for the asymmetries between jobs (please refer to the \textit{Supplementary Information} for a detailed review of the related literature).
Our system accounts for the asymmetries between occupations (it is easier to move in one direction than the other), leverages real-time job ads data at the granular skills-level, and accurately recommends occupations and skills that can assist workers looking to transition between jobs based on their personalized skills set.
We further demonstrate the flexibility of the \Method{} method by constructing a leading indicator of Artificial Intelligence (AI) adoption within Australian industries.
In our applications of \Method{}, we are able to both \textit{recommend} transition pathways to workers based on their personalized skill sets
and \textit{detect} emerging AI disruption that could accelerate job transitions.

\section*{Materials and methods}
\label{sec:mat-methods}

\subsubsection*{Datasets and ground-truth}

\paragraph{Job ads data.}
This research draws on 8,002,780 online job ads in Australia from 2012-01-01 to 2020-04-30, courtesy of Burning Glass Technologies (BGT).
This dataset provides unique insights into the evolving labor demands of Australia. It also covers the early periods of the COVID-19 crisis when Australian governments closed `non-essential' services~\cite{abc-lockdown}.
To construct this dataset, BGT has systematically collected job ads via web-scraping. 
This process removes duplicates of job ads posted across multiple job boards or job ads re-posted in short time-frames.
They also parse the unstructured job description text through their proprietary systems that extract key features from the advertised job.
These features include location, employer, salary, education requirements, experience demands, occupational class, industry classifications, among others. 
Importantly for this research, the skill requirements have also been extracted ($>$11,000 unique skills).
Here, BGT adopt a broad description of `skills' to include skills, knowledge, abilities, and tools \& technologies.
This is slightly different to the more commonly used skills data from O*NET, which defines skills as a series of developed capacities that are categorized into different competencies~\cite{US_Department_of_Labor_undated-oe}. 
There are two major advantages of using BGT job ads data over O*NET skills data: (1) more granular `skills' data and (2) longitudinal (when used historically) and near-real-time skills data in specific locations.
The latter point is particularly important when building a real-time job transitions recommender system to navigate labor crises as they unfold.

\paragraph{Employment statistics.}
The employment data used for this research is drawn from the `Quarterly Detailed Labor Force' statistics by the Australian Bureau of Statistics (ABS)~\cite{Australian_Bureau_of_Statistics2019-sv}.
These data represent labor supply features for the 4-digit occupations in the \textit{Job Transitions Recommender System} and include measures of employment levels and hours worked per occupation.

\paragraph{Occupational transitions ground-truth.}
The Household, Income and Labour Dynamics in Australia (HILDA) Survey is a nationally representative longitudinal panel study of Australian households that commenced in 2001~\cite{HILDA_2020}.
It has three main areas of interest: income, labor, and family dynamics. The HILDA survey is in its 18th consecutive year, with the latest available data available from 2018. 

Included within the HILDA  are data on occupational history and movements of anonymized respondents. We use this data to identify when respondents have changed jobs from one year to another. The occupations are recorded at the 4-digit level from the Australian and New Zealand Standard Classification of Occupations (ANZSCO). This shows the occupation of the previous year and the current year. We use this longitudinal dataset as the ground truth for validating \Method. As the job ads dataset used for this research begins at 2012, we isolate the observations of occupational transitions from 2012 to 2018 (the latest available year). This results in a sample of 2,999 occupational transitions in Australia.

\subsection*{Measuring skill similarity}
To measure the distance between occupations (or other skill groups), we first measure the pairwise distance between individual skills (6,981 skills in 2018) in jobs ads for each calendar year from 2012-2020.
Intuitively, two skills are similar when they are simultaneously important for the same set of job ads.
We measure the importance of a skill in a job ad using an established measure called `Revealed Comparative Advantage' ($RCA$ -- \cref{eq:rca}) that has been applied across a range of disciplines, such as trade economics~\cite{Hidalgo2007-qk,Vollrath1991-kr}, identifying key industries in nations~\cite{Shutters2016-fe}, detecting the labor polarization of workplace skills~\cite{Alabdulkareem2018-jl}, and adaptively selecting occupations according to their underlying skill demands~\cite{dawson2019adaptively}.
$RCA$ normalizes the total share of demand for a given skill across all job ads. 
We then calculate the pairwise skill similarity between each skill using \cref{eq:theta} as implemented by Alabdulkareem et al.~\cite{Alabdulkareem2018-jl} and again by Dawson et al.~\cite{dawson2019adaptively}.
These individual skill distances form the basis for measuring the distance between sets of skills.

To measure the distance between every skill for each year in the dataset, we start by removing extremely rare skills. 
Here, we select skills with a posting frequency count $>=5$, which represent $\sim 75\%$ of all skills (see \textit{Supplementary Information} for more details).
Let $\mathbb{S}$ be the set of all skills and $\mathbb{J}$ be the set of all job ads in our dataset.
We measure the similarity between two individual skills $s_1$ and $s_2$ ($s_1, s_2 \in \mathbb{S}$) using a methodology proposed by Alabdulkareem et al.~\cite{Alabdulkareem2018-jl}.
First, we use the \textit{Revealed Comparative Advantage} (RCA) to measure the importance of a skill $s$ for a particular job ad $j$:
\begin{equation} \label{eq:rca}
  RCA(j, s) = \frac{x(j, s) / \mathop{\sum}\limits_{s'\in \mathbb{S}}x(j, s')}
  {\mathop{\sum}\limits_{j'\in J}x(j', s) / \mathop{\sum}\limits_{j'\in \mathbb{J},s'\in \mathbb{S}}x(j', s')}
\end{equation}
\noindent
where $x(j,s) = 1$ when the skill $s$ is required for job $j$, and $x(j,s) = 0$ otherwise;
$RCA(j, s) \in \left[ 0, \mathop{\sum}\limits_{j'\in \mathbb{J},s'\in \mathbb{S}} x(j', s') \right], \forall j, s$, and the higher $RCA(j, s)$ the higher is the comparative advantage that $s$ is considered to have for $j$.
Visibly, $RCA(j, s)$ decreases when the skill $s$ is more ubiquitous (i.e. when $\mathop{\sum}\limits_{j'\in \mathbb{J}}x(j', s) $ increases), or when many other skills are required for the job $j$ (i.e. when $\mathop{\sum}\limits_{s'\in \mathbb{S}}x(j, s')$ increases).
Next, we measure the similarity between two skills based on the likelihood that they are both effectively used in the same job ads.
Formally:
\begin{equation} \label{eq:theta}
  \theta(s_1, s_2) = \frac{\mathop{\sum}\limits_{j \in \mathbb{J}}e(j,s_1)\cdot e(j,s_2)}
  {max \left( \mathop{\sum}\limits_{j\in \mathbb{J}}e(j,s_1), \mathop{\sum}\limits_{j\in \mathbb{J}}e(j,s_2) \right)}
\end{equation}
where $e(j, s)$ is the effective use of a skill in a job, defined as $e(j, s) = 1 \text{ when } RCA(j,s) \ge 1$ and $e(j, s) = 0$ otherwise.
Note that $\theta(s_1, s_2) \in [0, 1]$, a larger value indicates that $s_1$ and $s_2$ are more similar, and it reaches the maximum value when $s_1$ and $s_2$ always co-occur (i.e. they never appear separately) while $e(j, s_1) = 1$ and $e(j, s_2) = 1, \forall j \in \mathbb{J}$.
Visibly, $\theta(s_1, s_2)$ is based on the co-occurrence of skills when both $s_1$ and $s_2$ are simultaneously important for the job ads.
Therefore, $\theta$ measures when two skills are effectively used together -- i.e., it measures similarity as in ``complementary'', not as in ``replaceable''.

\subsection*{\Method Method}
Next, we use the pairwise skill distances to measure the distance between \textit{sets of skills}, which we refer to as \Method. 
Here, a set of skills can be arbitrarily defined, such as an occupation, an industry, or a personalized skills set. 
Intuitively, two sets of skills are similar when their most important skills are similar.
We first introduce a measure of skill importance within a skill set as the mean RCA over all the job ads pertaining to the skill set. 
Assume a job ads grouping criterion exists, for example, job ads pertaining to an occupation, a company, or an industry. 
We obtain the job ads set $\mathcal{J} \subset \mathbb{J}$ and the set of skills $\mathcal{S} \subset \mathbb{S}$ occurring within $j \in \mathcal{J}$.
We denote $\mathcal{J}$ as the set of job ads associated with the skill set $\mathcal{S}$.
We measure the importance of skill $s$ for $\mathcal{S}$ (and implicitly for $\mathcal{J}$) as the mean RCA over all the job ads relating to the skill set $\mathcal{S}$. 
Formally,
\begin{equation} \label{eq:skill-importance}
  w(s, \mathcal{S}) = \frac{1}{|\mathcal{J}|} \mathop{\sum}\limits_{j \in \mathcal{J}} RCA(j, s)
\end{equation}

Next, we propose a method to measure the distance between \textit{sets of skills}.
For example, suppose there are two jobs that we can define by their underlying skill demands. 
Both jobs have their unique set of skills, and each individual skill has its own relative importance to the specific job, as calculated by \cref{eq:skill-importance}.
Intuitively, the two jobs are similar when their most important skills (i.e., their `core' skills) are similar.
This is achieved by computing the weighted pairwise skill similarity between the individual skills of each job (using \cref{eq:theta}), where the weights correspond to the skill importance (defined by \cref{eq:skill-importance}).
This returns a single similarity score between the two skill sets corresponding to the two jobs.
Formally, let $\mathcal{S}_1$ and $\mathcal{S}_2$ be two sets of skills, and $\mathcal{J}_1$ and $\mathcal{J}_2$ their corresponding sets of job ads.
We define $\Theta$ the similarity between $\mathcal{S}_1$ and $\mathcal{S}_2$ as the weighted average similarity between the individual skills in each set, where the weights correspond to the skill importance in their respective sets.
Formally,
\begin{equation} \label{eq:skill-set-sim}
  \Theta(\mathcal{S}_1, \mathcal{S}_2) = \frac{1}{C} \mathop{\sum}\limits_{s_1 \in \mathcal{S}_1} \mathop{\sum}\limits_{s_2 \in \mathcal{S}_2} w(s_1, \mathcal{S}_1) w(s_2, \mathcal{S}_2) \theta(s_1, s_2)
\end{equation}
where $C = \mathop{\sum}\limits_{s_1 \in \mathcal{S}_1} \mathop{\sum}\limits_{s_2 \in \mathcal{S}_2} w(s_1, \mathcal{S}_1) w(s_2, \mathcal{S}_2)$.
Similar to $\theta$ defined in \cref{eq:theta}, $\Theta$ is a similarity measure (higher means more similar) and $\Theta(\mathcal{S}_1, \mathcal{S}_2) \in [0, 1]$. 
Note that $\Theta$ is a compound measure based on $\theta$, which in turn measures the complementarity of two skills (see prior discussion and interpretation of $\theta$).
As a result, $\Theta(\mathcal{S}_1, \mathcal{S}_2)$ measures the complementarity of two skill sets. 
The interpretation we use in the rest of this paper is that ``when $\Theta$ is high, an individual with $\mathcal{S}_1$ can more readily fulfil the skill requirements of $\mathcal{S}_2$''.
We use $\Theta$ as a key feature in our job transitions recommender system. 
The setup and details of this system now follow.

\subsection*{Job Transitions Recommender System setup} \label{subsec:job-transitions-setup}
We construct the job transitions recommender system as a binary machine learning classifier using XGBoost -- an implementation of gradient boosted tree algorithms, which has achieved state-of-the-art results on many standard classification benchmarks with medium-sized datasets~\cite{XGBoost}.
The XGBoost algorithm is a linear combination of decision trees where each subsequent tree attempts to reduce the errors from its predecessor.
This allows the next tree in the series to `learn' from the errors of the previous tree, with the goal of making more accurate predictions.
In our application of the XGBoost algorithm, the system `learns' from the input labor market data, which are independent variables (or features).
It is then `trained' against historical examples of occupational transitions that did occur (positive examples) and did not occur (negative examples), which are the dependent variables (or ground-truth).
As is standard in machine learning practice, we reserve a `test set' of observations for evaluation, where we apply the trained model to make predictions about whether a transition occurred or not (hence, binary) by only observing the features.
This setup allows us to predict the probability of an occupational transition from the `source' to the `target' occurring (positive example) or not (negative).
Here, we use the job-to-job transitions data from the HILDA dataset~\cite{HILDA_2020} (described above). 
We then randomly simulate an alternate sample of transitions where we maintain the same `source' occupations and randomly select `target' occupations (called `Random Sample').
This produces a balanced dataset of 5,998 positive and negative occupational transitions. 
We then associate each `source' and `target' observation with their temporal pairwise distance measure using the \Method method. 
However, the \Method measures are symmetric, and job transitions are known to be asymmetric~\cite{robinson2018occupational, nedelkoska2015skill, MovingBetweenJobs}. 
Therefore, to represent the asymmetries between job transitions, we add a range of explanatory features to each `source' and `target' occupation.
These occupational features include their \Method pairwise distance measures and other variables, such as years of education required, years of experience demanded, and salary levels, from employment statistics (`Labor Supply') and job ads data (`Labor Demand' -- see \textit{Supplementary Information} for full list of features).

Like most machine learning algorithms, XGBoost has a set of hyper-parameters -- parameters related to the internal design of the algorithm that cannot be fit from the training data. 
The hyper-parameters are usually tuned through search and cross-validation.
In this work, we employ a Randomized-Search~\cite{bergstra2012random} which randomly selects a (small) number of hyper-parameter configurations and performs evaluation on the training set via cross-validation.
We tune the hyper-parameters for each learning fold using 2500 random combinations, evaluated using a 5 cross-validation. 
We train the models on 80\% of the observations, leaving aside 20\% of the data for testing, which we randomly seed.
We repeat the process 10 times for each feature model configuration and change the random seed to select a new testing sample, which provides us with the standard deviation bars seen in \cref{fig:validation-trans-map}. 

\subsection*{Constructing a leading indicator of AI adoption}
Adapting the \Method method, we develop a leading indicator for emerging technology adoption and potential labor market disruptions based on skills data, using AI as an example. 
We select AI because of its potential impacts on transforming labor tasks and accelerating job transitions~\cite{brynjolfsson2014second, Frey2017, Frank2019-dn}.
Our indicator temporally measures the similarity between a dynamic set of a AI skills against the 19 Australian industry skill sets from 2013-2019.

To create these yearly sets of top AI skills, we first select a sample of core `seed skills' that are highly likely to remain important to AI over time -- here we selected `Artificial Intelligence', `Machine Learning', `Data Science', `Data Mining', and `Big Data' as the seed skills. 
This set of seed skills represents $\mathcal{S}$ from \cref{eq:skill-importance} as opposed to a grouping criterion, such as an occupation or industry. 
In this case, $\mathcal{J}$ is not defined, and we measure the importance of a skill as the average $\theta$ similarity to the seed skills.
Repeating this process temporally allows us to build dynamic skill sets.
We then use \cref{eq:theta} to measure the similarity ($\theta$) of each seed skill to every other skill in a given year. 
By calculating the average $\theta$ for every skill relative to the seed skills, we return an ordered list of skills with the highest similarity.
This process is repeated for each calendar year from 2013-2019 where we select the top 100 skills for each year.
The skill similarity approach allows us to build an adaptive list of AI skills that captures evolving skill demands.
This is especially important for a skill area like AI, where the skill demands are changing very quickly.
For example, `TensorFlow' (a Deep Learning framework) emerged as a skill in November 2015 and has since become among the fastest-growing AI skills. 
The AI skill lists we create can detect the importance of `TensorFlow' in 2016, whereas a static list pre-defined before 2016 would have missed these important changes to AI skill demands. 

Having constructed temporal sets of AI skills, we then measure the yearly similarities between the AI skillsets and the skill sets of Australia's 19 major industries -- classified according to the Australian \& New Zealand Standard Industrial Classification (ANZSIC) Division level.
Using the \Method method, we construct each industry as a set of skills for every year and use \cref{eq:skill-set-sim} to calculate similarity to the yearly AI skill sets.
This allows us to observe and compare the extent to which AI skills have diffused throughout industries and the relative importance of AI skills to these industries.
The advantages of using this skill similarity approach as opposed to \textit{ad hoc} skill counts from pre-defined skills are twofold.
First, we create dynamic sets of skills that capture evolving skill demands.
Second, we account for skill importance within individual job ads by normalizing for high-occurring skills (see \textit{Supplementary Information} for more details).

\section*{Results \& Discussion}
\subsection*{Skill similarity results}
\begin{figure}[]
	\centering

\includegraphics[height = 0.74\textheight]{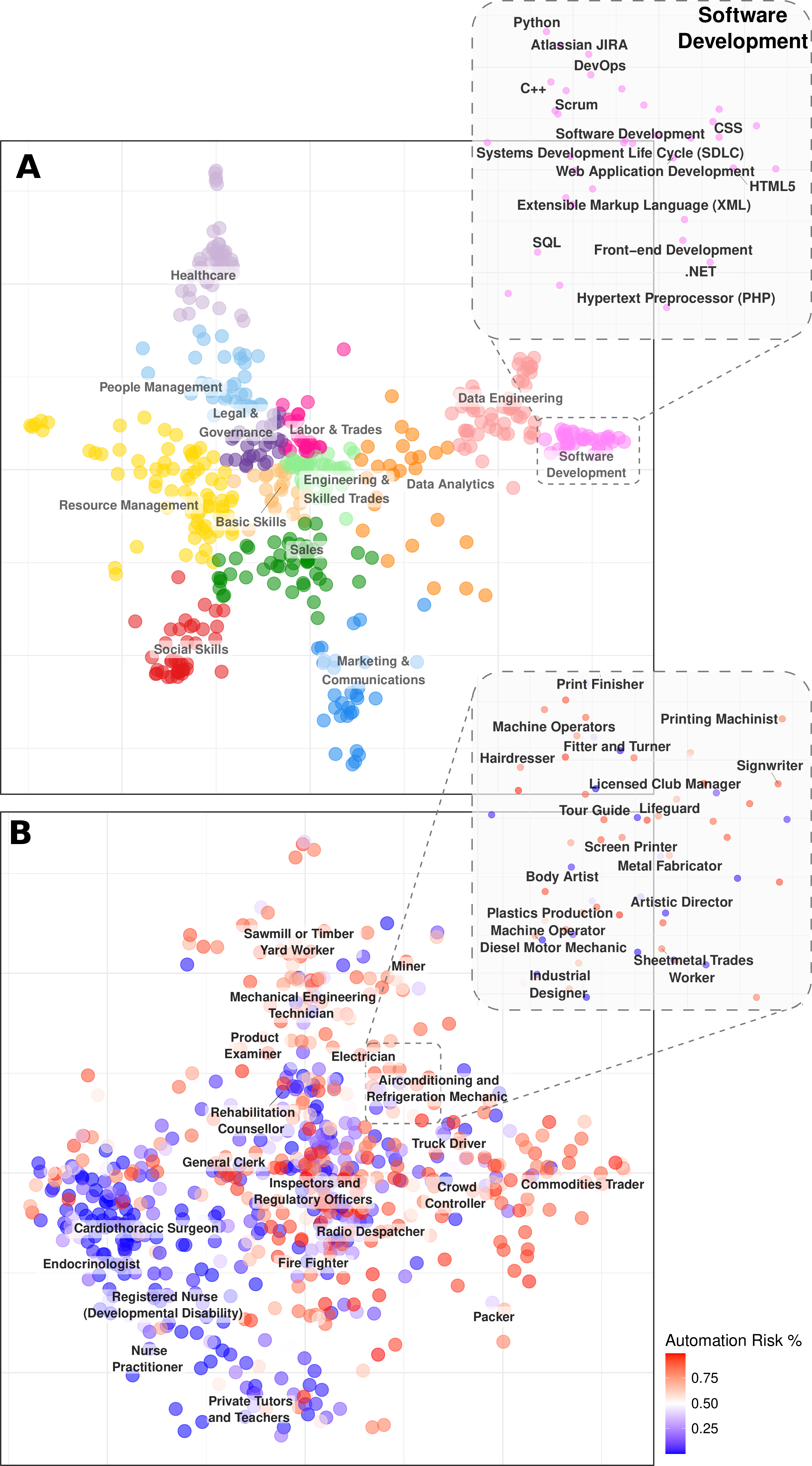}

	\caption[]{
		\textbf{Measuring the distance between skills and occupations. }
		(A) Shows a two-dimensional embedding of the 2018 skill distances, where the top 500 skills by posting frequency are visualized. Each marker represents an individual skill colored according to 13 clusters using K-Medoids clustering. As observed in the `Software Development' inset, highly similar skills cluster together. Additionally, the specialized skill clusters, such as `Software Development' and `Healthcare' tend to lay toward the edges; whereas the more general and transferable skills lay toward the middle of the embedding and act as a `bridge' to specialist skills. These individual skill distances form the basis of measuring the distance between sets of skills. (B) We leverage \Method to measure the distance between official Australian occupations at the 6-digit level (characterized by their skill sets) in 2018. The markers represent individual occupations, colored by technological labor automation risk, as calculated by Frey \& Osborne. Occupations that require higher levels of routine, manual and/or low cognitive labor tasks tend to be at higher risk (colored darker red); whereas occupations characterised by non-routine, interpersonal, and/or high cognitive labor tasks are at lower risk (colored darker blue) over the next two decades.
	}
	\label{fig:similarity-maps}
\end{figure}

\cref{fig:similarity-maps}A shows the two-dimensional skill distance embeddings for the top 500 skills by posting frequency in 2018. 
Here, each marker represents an individual skill that is colored according to one of 13 clusters of highly similar skills, using the K-Medoids clustering algorithm.
By using a triplets method for dimensionality reduction~\cite{2019TRIMAP}, we are able to preserve the global structure of the embedding (global structure = 98\%).
That is, two markers are depicted closer together when their corresponding skills are more similar (i.e., have low distance).
This provides useful insights, highlighting that specialized skills (such as `Software Development' and `Healthcare') tend to lay toward the edges of the embedding, whereas more general and transferable skills lay toward the middle, acting as a `bridge' to specialist skills.
Highly similar skills cluster closely together; for example, the `Software Development' cluster (see inset) regroups programming skills such as scripting languages `Python', `C++', and `HTML5'. 
It is important to measure the similarity between jobs based on their underlying skills because workers leverage their existing skills to make career changes~\cite{brynjolfsson2013complementarity}.

\subsection*{\Method results}
In \cref{fig:similarity-maps}B, the markers depict groups of skills that correspond to individual occupations,
using the official Australian standard (at the 6 digit level -- see \textit{Supplementary Information} for more details). 
To visualize the distance between occupations, we use the same dimensionality reduction technique as for individual skills in \cref{fig:similarity-maps}A.
Each occupation is colored on a scale according to their automation susceptibility, as calculated by Frey and Osborne~\cite{Frey2017} -- dark blue represents low-risk probability, and dark red shows high-risk probability over the coming two decades. 
As seen in \cref{fig:similarity-maps}B and the magnified inset, similar occupations lie close together on the map.
Furthermore, occupations at low risk of automation tend to be characterized by non-routine, interpersonal, and/or high cognitive labor tasks~\cite{autor2013update}. 
In contrast, occupations at high risk of automation require routine, manual, and/or low cognitive labor tasks.
For example, in the inset of \cref{fig:similarity-maps}B, a `Sheetmetal Trades Worker' is deemed to be at high risk of labor automation (82\% probability) due to high levels of routine and manual labor tasks required by the occupation. 
However, a `Sheetmetal Trades Worker' skillset demands are highly similar to an `Industrial Designer', which is considered at low risk of labor automation over the coming two decades (4\% probability). 
Therefore, an `Industrial Designer' represents a transition opportunity for a `Sheetmetal Trades Worker' that leverages existing skills and protects against potential risks of technological labor automation.

\begin{figure}[]
    \centering
\includegraphics[width=.95\linewidth]{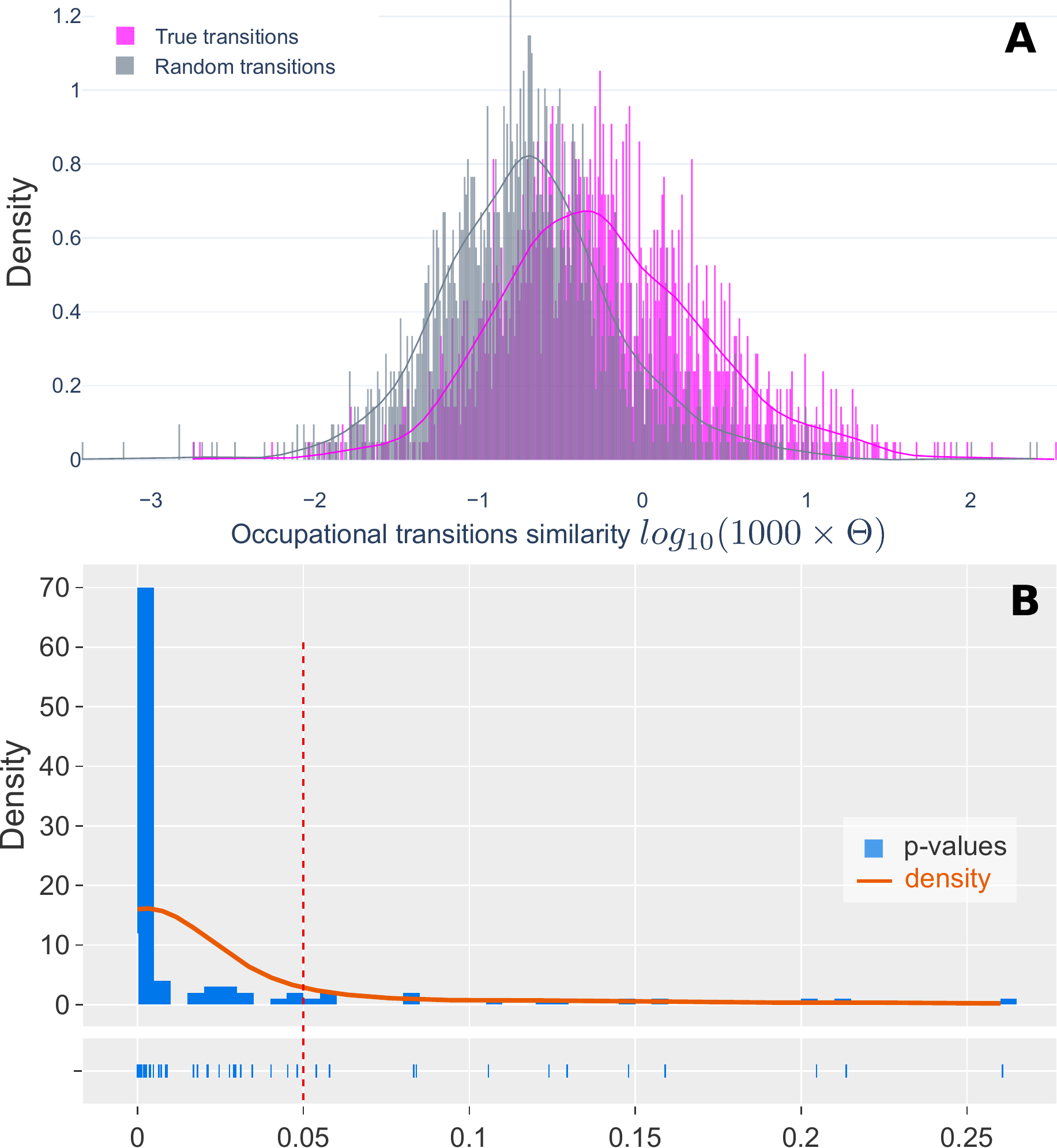}
    \caption{
		\textbf{The \Method is statistically significant in representing job transitions.}
		\textbf{(A)} The x-axis shows the log-transformed \Method distance for a `True' sample of actual transitions (shown in magenta color) and a `Random' sample of simulated transitions (shown in gray color).
		Each Random transition is paired with an Actual transition: it shares the same `source' occupation as the Actual transition but the `target' occupation is randomly selected and is different to the Actual observation.
		The difference between the two populations is statistically significant (paired t-test, t-statistic = 4.514, p-value = $6.535 \times 10^{-6}$, Cohen's D effect size = 0.14).        
\textbf{(B)} We repeat the procedure 100 times: we generate 100 `Random' populations, and we perform the statistical testing for each.
		The figure shows the histogram (density and rug plot) of the 100 obtained p-values, 87 of which are lower than 0.05.
	}
    \label{fig:theta-validation}
\end{figure}

\clearpage
\subsection*{Validation of \Method distance}
\label{subsec:stat-test}
We validate the link between the \Method and job transitions by conducting paired t-tests, as illustrated in \cref{fig:theta-validation}.
Here, we use a longitudinal household survey dataset containing actual job transitions~\cite{HILDA_2020} (called the `True Sample').
Each occupational pair (`source' to `target') is labeled with its \Method distance for the given year.
We randomly simulate an alternate transition sample by maintaining the same `source' occupations and randomly selecting `target' occupations (called `Random Sample').
Our objective is to determine whether the differences in \Method distance between the `True Sample' and the `Random Sample' are statistically significant.
First, we test the differences of the two samples, including \textit{all} occupational transitions. 
We find that the differences between the two samples are statistically significant (t-statistic = $16.272$, p-value = $2.707 \times 10^{-58}$, Cohen's D effect size = $0.42$) (see \textit{Supplementary Information}). 
However, one-third of the `True Sample' transitions are to another job but are classified as the same occupation.
Therefore, we perform a second test only on transitions between different occupations, i.e., we remove all observations where the `source' and `target' are identical. 
Again, the differences between the `True Sample' and the `Random Sample' are statistically significant (t-statistic = $4.514$, p-value = $6.535 \times 10^{-6}$, Cohen's D effect size = $0.14$), as illustrated in \cref{fig:theta-validation}A.
We repeat the procedure 100 times: we generate 100 new `Random Samples', and we perform the statistical test for each of them.
$87\%$ of these tests are statistically significant, as shown in \cref{fig:theta-validation}B.
These results provide evidence that the \Method distance measure is representative of actual job transitions.

\subsection*{Job Transitions Recommender System}

\begin{figure*}[]
	\centering
\includegraphics[width=1\linewidth]{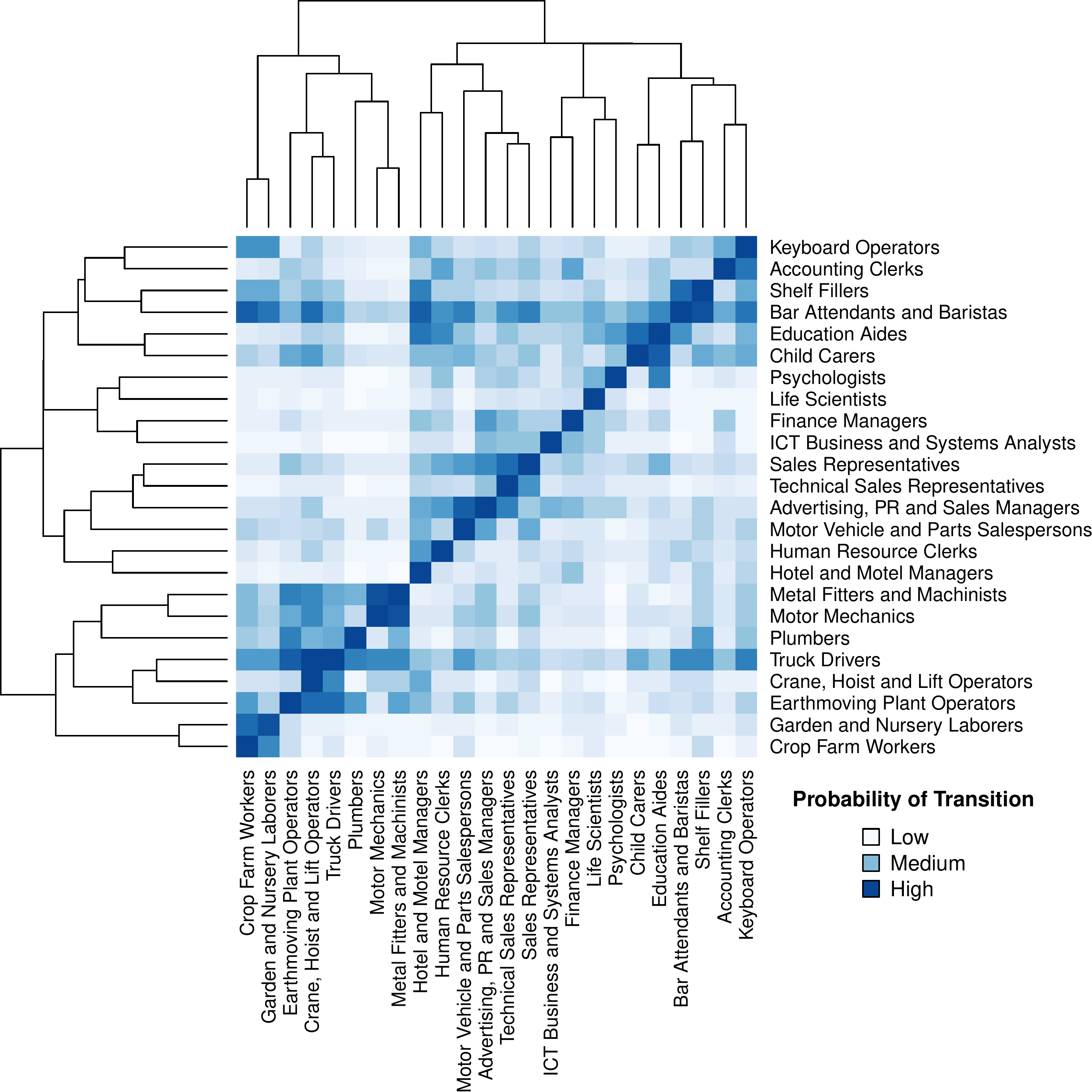}
	\caption{
    Validation and the \textit{Transitions Map}.
		Visualizes a subset of the \textit{Transitions Map}, where 20 occupations and their pairwise transition probabilities can be observed.
		In this visualization, transitions occur from columns to rows, and dark blue depicts high transition probabilities, and white depicts low probabilities. While job transitions to the same occupation yield the highest probabilities (dark blue diagonal squares), it is clear that transitions are asymmetric.
		The dendrogram highlights how similar occupations cluster together, where there is a clear divide between services and manual labor occupations.
	}
	\label{fig:validation-trans-map}
\end{figure*}

Job transitions, however, are \textit{asymmetric}~\cite{robinson2018occupational, nedelkoska2015skill, MovingBetweenJobs} -- the direction of the transition affects the difficulty.
Therefore, transitions are determined by more than the symmetric distance between skill sets; other factors, such as educational requirements and experience demands, contribute to these asymmetries.
We account for the asymmetries between job transitions by constructing a machine learning classifier framework that combines the \Method distance measures with other labor market features from job ads data and employment statistics (discussed in \textit{Job Transitions Recommender System setup}). 
We then apply the trained model to predict the probability for every possible occupational transition in the dataset -- described as the transition probability between a `source' and a `target' occupation.
This creates the \textit{Transitions Map}, for which a subset of 20 occupations can be seen in \cref{fig:validation-trans-map}. 
The colored heatmap shows the transition probabilities (`source' occupations are in columns and `targets' are in rows). Dark blue represents higher transition probabilities, and lighter blue shows lower probabilities, where the asymmetries between occupation pairs are clearly observed.
For example, a `Finance Manager' has a higher probability of transitioning to become an `Accounting Clerk' than the reverse direction. 
Moreover, transitioning to some occupations is generally easier (for example, `Bar Attendants and Baristas') than others (`Life Scientists'). 
The dendrogram illustrates the hierarchical clusters of occupations where there is a clear divide in \cref{fig:validation-trans-map} between service-oriented professions and manual labor occupations.

\begin{figure}[t]
    \centering
\includegraphics[width=.75\linewidth]{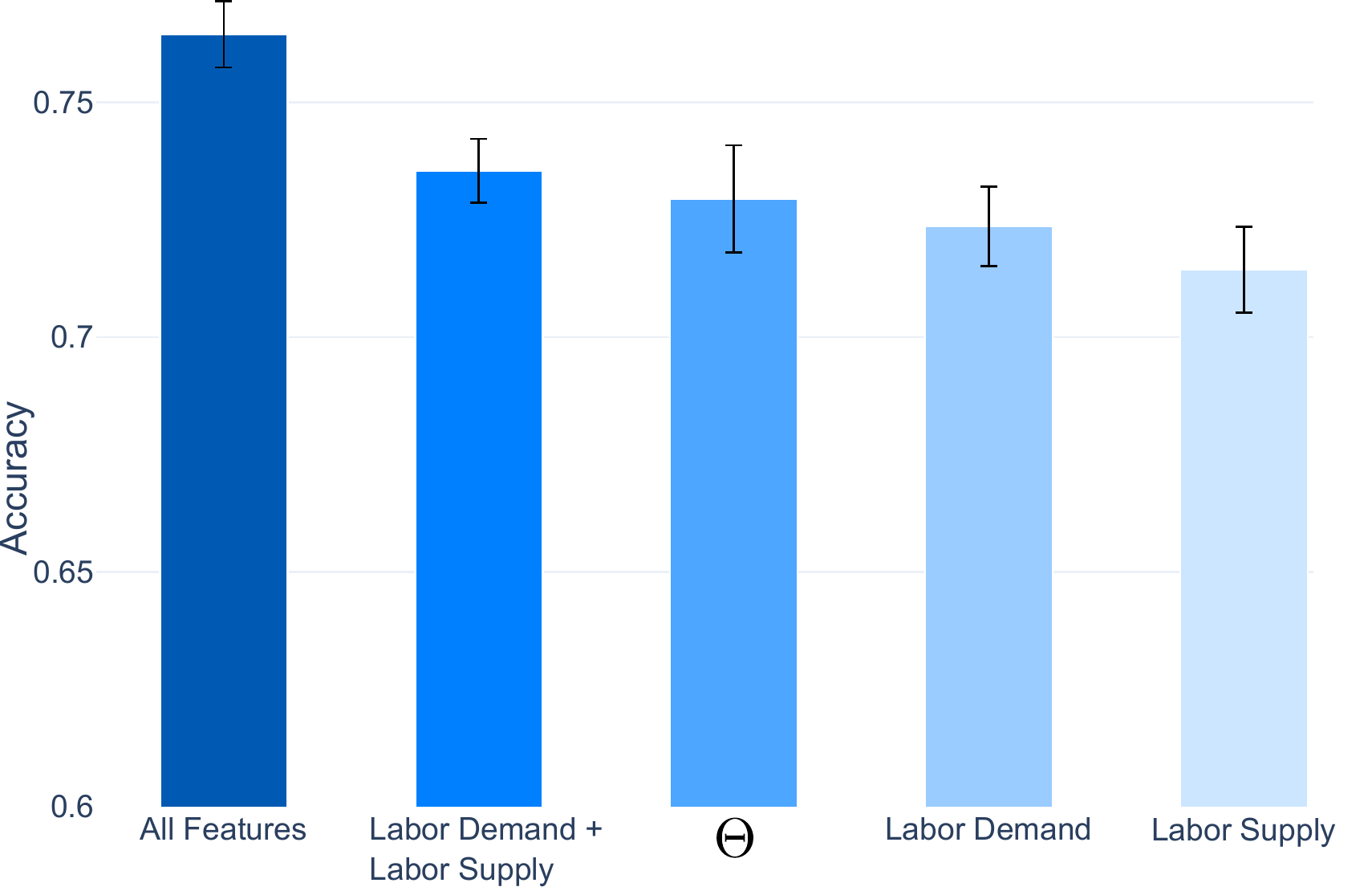}
    \caption{
        The prediction accuracy scores of the different classifier model feature configurations.
		The highest performance is achieved when `All Features' are incorporated in the classifier models to predict occupational transitions (76\%). 
		Moreover, by incorporating all features, the standard deviation decreases (shown by the performance bars),
		which highlights the complementarity of the combined features and the ability to now account for the asymmetry between jobs.}
	\label{subfig:accuracy}
\end{figure}

For validation, we train various classifier models with different feature configurations and identify three main findings.
\textit{First}, as seen in \cref{subfig:accuracy}, the models that incorporate the distance measures with all of the other occupational features (`All Features') consistently achieve the highest accuracy for occupational transitions (Accuracy = 76\%).
This feature setup achieves higher results than models that only use the `Labor Demand + Labor Supply' features (Accuracy = 74\%) or the distance measures alone (Accuracy = 73\%). 
To further validate these findings, we conduct an ablation test where each feature is iteratively removed from the feature set and the model is re-trained -- the model configurations with lower performance indicate higher feature importance. 
Here, the exclusion of the \Method{} distance measure caused the greatest decline in performance, therefore reiterating its predictive power. 
We also conduct a feature importance analysis, which again shows that the \Method{} distance measure is the most important feature for predicting transitions (see \textit{Supplementary Information} for further details).
\textit{Second}, the standard deviation of accuracy over repeated trials decreases when all features are combined (as seen with the spread of the performance bars in \cref{subfig:accuracy}). This shows that the \Method{} distance measures and the occupational features are complementary in predicting job transitions.
\textit{Third}, and most important, is that by combining all features, we can construct the asymmetries between occupations. 
While transitioning to a job in the same occupation yields the highest probabilities (as seen by the dark blue diagonal line in \cref{fig:validation-trans-map}), the occupational features add asymmetries between occupational pairs, such as accounting for asymmetries in education and experience requirements.

\subsubsection*{Recommending Jobs and Skills.}
\label{sec:recommending-jobs}

\begin{figure}\centering
\includegraphics[width=0.95\linewidth]{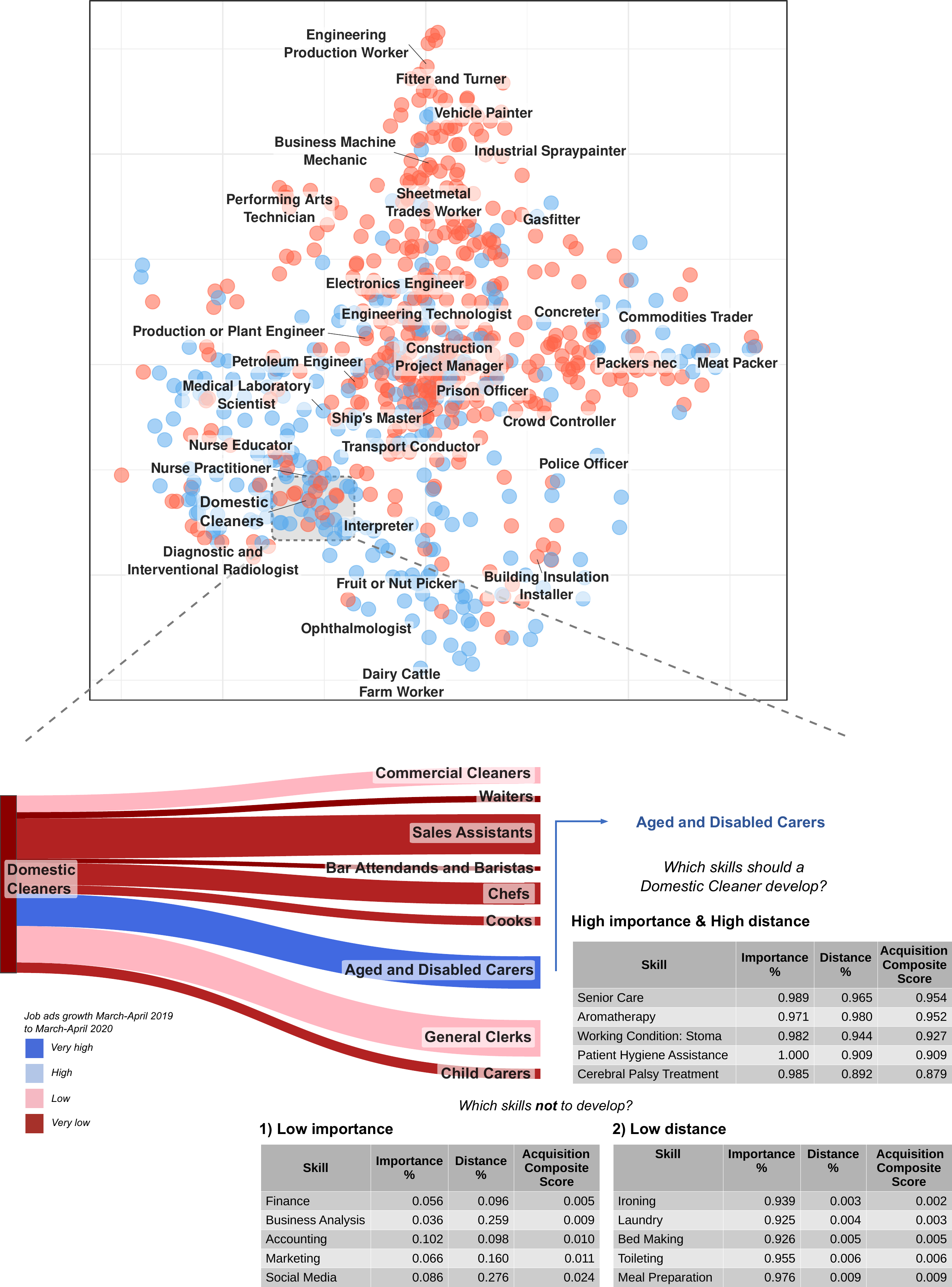}
	\captionsetup{font=footnotesize,labelfont=footnotesize}
	\caption{Here, we demonstrate the \textit{Job Transitions Recommender System} using the `Domestic Cleaner' occupation as an example -- an occupation classified as a `non-essential' occupation and that has experienced significant declines during the beginning of the COVID-19 outbreak in Australia.
  (upper panel) The two-dimensional space of occupations (see \mbox{\cref{fig:similarity-maps}}) with `essential' occupations in blue markers and `non-essential' occupations in red markers.
  (lower panel)
  We first use the \textit{Transitions Map} to calculate the occupations with the highest transition probabilities (other than itself). These are the nodes on the right side of the flow diagram in the bottom panel of the figure, where the link colors depict posting frequency percentage change from March-April 2019 to March-April 2020. The link widths represent the posting frequency volume of March-April 2020 to indicate labor demand. The top six occupations have all experienced significant declines during the COVID-19 period; however, the seventh recommendation, `Aged and Disabled Carers',  is experiencing significant labor demand growth. `Aged and Disabled Carers' were also classified as an `essential' occupation during COVID-19 in Australia. We select this as the target occupation and then make personalized skill recommendations. We argue that workers trying to transition to another occupation should invest time and resources into skill development when (1) the skills are of \textbf{high importance} \textit{and} (2) there is a \textbf{high distance} to acquire the skill. Conversely, workers should \textit{not} focus on skill development if (1) the skills are \textbf{low importance} \textit{or} (2) there is a \textbf{low distance} to acquire the skill.
  }
	\label{fig:transition-example}
\end{figure}

The \textit{Transitions Map} provides the basis for making qualified job transition recommendations. 
We call this the \textit{Job Transitions Recommender System}.
In \cref{fig:transition-example}, we showcase its usage in the context of a labor market crisis (i.e. COVID-19). 
We highlight the example of `Domestic Cleaners', an occupation that experienced significant declines in labor demand and employment levels during the crisis (see \textit{Supplementary Information}). 

During the `first wave' of COVID-19 in March 2020, the Australian Government enforced social-contact and mobility restrictions on `non-essential services' to slow the spread of the virus~\cite{Harris2020-dt}. 
Many occupations within these `non-essential' services were unable to trade and perform their duties, forcing some workers to try and transition to another job. 
In the upper panel of \cref{fig:transition-example}, we visualize the `essential' and `non-essential' occupations on the \textit{Transitions Map}.
The placement of the occupations is identical to \cref{fig:similarity-maps}B and we highlight the `essential' occupations as the blue nodes and the `non-essential' occupations are the red nodes using the classifications developed by Faethm AI~\cite{Faethm_data}.
We observe that the cluster of medical occupations (bottom-left of the map) are deemed as `essential', as are most of the food production workers (bottom).
Here, we select `Domestic Cleaners' as an example of a `non-essential' occupation and use the \textit{Transitions Map} to recommend the occupations with the highest transition probabilities in the bottom panel of \cref{fig:transition-example}.
These are the nodes on the right side of the flow diagram in \cref{fig:transition-example}, ordered in descending order of transition probability. 
Segment widths show the labor demand for each of the recommended occupations during the COVID-19 period (measured by posting frequency).
The segment colors represent the percentage change of posting frequency during March and April 2019 compared to the same months in 2020; dark red indicates a big decrease in job ad posts, and dark blue indicates a big increase. 
The first six transition recommendations for `Domestic Cleaners' have all experienced negative demand, which is unsurprising given that they were also deemed as `non-essential' services.
However, the seventh recommendation, `Aged and Disabled Carers', has significantly grown in demand during the COVID-19 period, and there is a high number of jobs advertised.
`Aged and Disabled Carers' is both an `essential' and a high-demand occupation; therefore, it makes sense to select `Domestic Cleaner' as the target occupation for transitioning into.

We take the \textit{Job Transitions Recommender System} a step further by incorporating skill recommendations. 
Transitioning to a new occupation generally requires developing new skills under time and resource constraints. 
Therefore, workers must prioritize which skills to develop. 
We argue that a worker should invest in developing a skill when (1) the skill is important to the target occupation \textit{and} (2) the distance to acquire the skill is large (that is, it is relatively difficult to acquire).
Formally, for a target skill (i.e., a skill in the `target' occupation), we compute its \emph{importance} to the target occupation and its \emph{distance} to the source occupation.
We calculate \emph{skill importance} as the mean $RCA$ for the skill across all job ads within the target occupation using \cref{eq:skill-importance}.
We calculate \emph{skill distance} as the distance between the target skill and `source' occupation skill set as $1 - \Theta(S_1, S_2)$ using \cref{eq:skill-set-sim} (i.e., the `target' skillset ($S_2$) is made out of a single skill: the target skill).
Finally, we build the \emph{acquisition composite score} as the multiplication of importance and distance, transformed as percentiles to account for variation in scale.

In the case of the `Domestic Cleaner' in \cref{fig:transition-example} (lower panel), the skills with the highest acquisition composite score for the transition to `Aged and Disabled Carer' are specialized patient care skills, such as `Patient Hygiene Assistance'. 
Conversely, the reasons not to develop a skill are when (1) the skill is not important \textit{or} (2) the distance is small to the target occupation. 
\cref{fig:transition-example} (lower panel) shows that while some `Aged and Disabled Carer' jobs require `Business Analysis' and `Finance' skills, these skills are of low importance for the `Aged and Disabled Carer' occupation, so they should not be prioritized. 
Similarly, skills such as `Ironing' and `Laundry' are required by `Aged and Disabled Carer' jobs, but the distance is small, so it is likely that either a `Domestic Cleaner' already possesses these skills or they can easily acquire them. 
Visibly, for both of the latter cases, the acquisition composite score takes low values.

\subsection*{A leading indicator of AI adoption} \label{sec:ai-adoption}
Emerging technologies can change the demand for labor and accelerate forced job transitions by disrupting labor markets~\cite{bresnahan2002information}. 
However, in order for emerging technologies to have these effects, they must first be widely adopted by firms across many industries. 
In this sense, technology adoption rates are a precursor to the societal impacts that they impose, such as widespread changes to labor demand and accelerated job transitions.
Measuring technology adoption, however, can be difficult as it often depends on the private activities of firms and is influenced by a range of factors (see \textit{Supplementary Information}).
Therefore, measuring the drivers that enable emerging technology adoption (see \textit{Supplementary Information}) can provide leading indicators of adoption rates.
One major driver is the availability of skilled labor.
Firms that can readily access workers with relevant skills are able to make productive use of the emerging technologies and accelerate their adoption rates, particularly in the early stages of technology growth~\cite{Bessen2015-pp}.
\Method offers a useful method for identifying the extent of specific skill gaps of firms within industries. 
As an industry's skills set becomes more similar to the skills of an emerging technology, the skills gap is narrowed, and the barriers to adopting the emerging technology are reduced.
When access to relevant skilled labor is plentiful, the labor requirements enabling technological adoption can be readily achieved and help accelerate adoption rates. 
Therefore, measuring temporal levels of skill set similarity for an emerging technology within an industry provides a useful leading indicator of technology adoption over time.
These measures offer early detection signals of emerging technology adoption and the changing skill requirements that could cause labor disruptions within industries, such as forced job transitions. 

\begin{figure}
\centering
\includegraphics[width=1\textwidth]{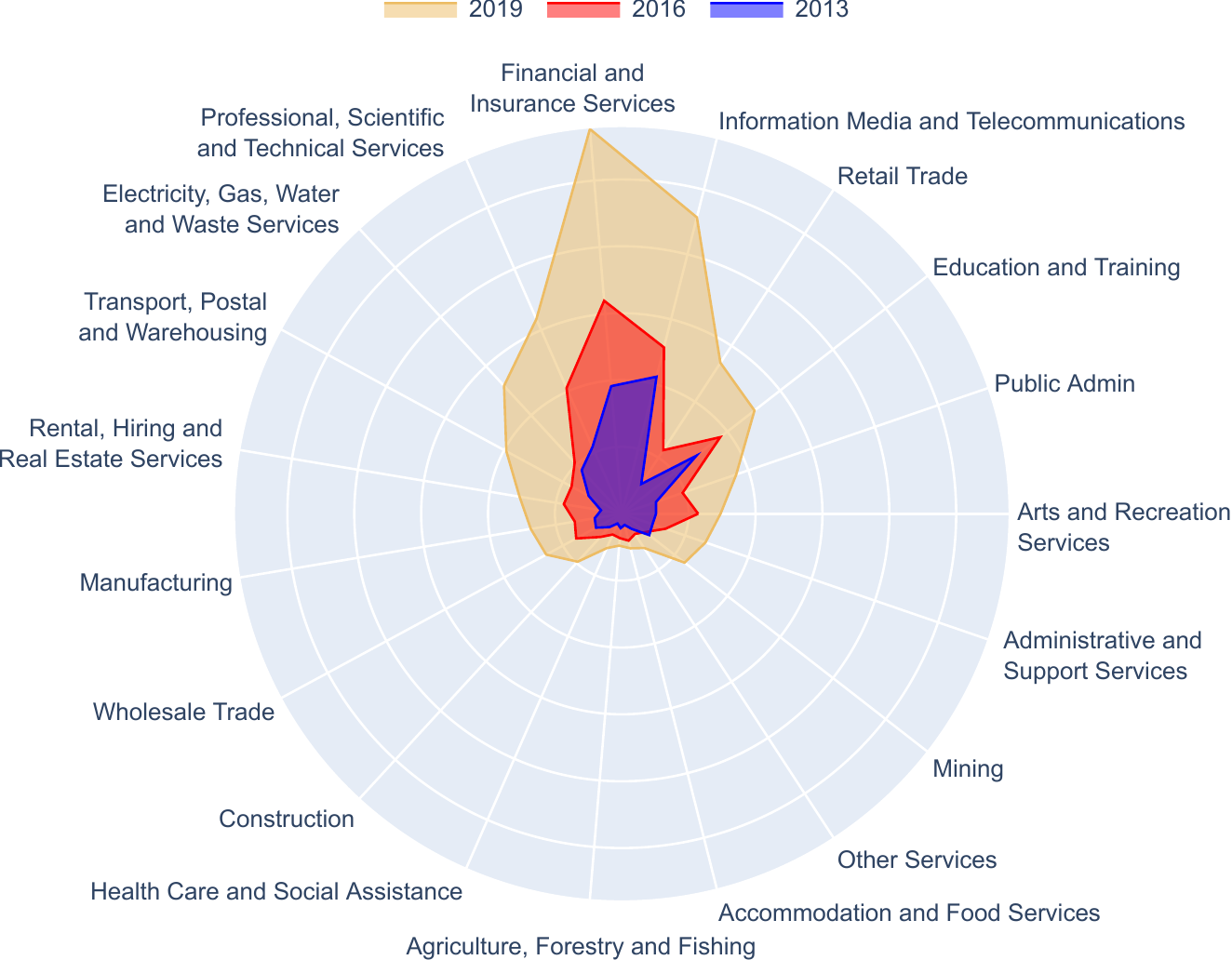}
\caption{By applying \Method, we measure the yearly similarities between adaptive sets of AI skills against each of the 19 Australian industries from 2013-2019. 
As industry skill sets become more similar to AI skills, the colored area of the radar chart expands. 
All industries have increased their similarity levels to AI skills, albeit at different rates. 
We argue that higher levels of AI similarity indicate AI skills are becoming more important to firms within an industry and that the skills gap to acquiring AI skills is narrowed. 
Access to these skills accelerates the rate of firms adopting AI and making productive use of the technologies, which offers a leading indicator of AI adoption and potential labor disruptions within these industries.}
\label{fig:ai-radar}
\end{figure}

\cref{fig:ai-radar} shows that all Australian industry skill sets have grown in similarity to AI skills from 2013 to 2019 -- illustrated by the expanding colored areas. This highlights the growing importance of AI skills across the Australian labor market. However, the rates of similarity are unequally distributed. 
Some industries -- such as `Finance and Insurance Services' and `Information Media and Telecommunications' -- command much higher rates of AI skill similarity. 
This indicates that not only are firms within these industries increasingly demanding AI skills but also that the AI skills gaps within these industries are much smaller.
Also noteworthy are the differences in growth rates toward AI skill similarity.
As clearly seen in \cref{fig:ai-radar}, AI skill similarity has rapidly grown for some industries and more modestly for others. For instance, `Retail Trade' has experienced the highest levels of growth in similarity to AI skills, increasing by 407\% from 2013 to 2019. The majority of this growth has occurred recently, which coincides with the launch of Amazon Australia in 2017~\cite{Amazon2019}. 
Since then, Amazon has swiftly hired thousands in Australia.

By adapting the \Method method, we develop a leading indicator that detects AI adoption from real-time job ads data.
Such a measure can act as an `early warning' signal of forthcoming labor market disruptions and accelerated job transitions caused by the growth of AI.
This indicator can assist policy-makers and businesses to robustly monitor the growth of AI skills (or other emerging technologies), which acts as a proxy for AI adoption within industries (or other labor market groups).

\subsection*{Limitations}\label{subsec:limitations}
We acknowledge several limitations of the \Method method and the results presented in this paper. 
First, there are \textit{data limitations} from both the household survey data and the job ads data. 
The job transitions drawn from the HILDA panel dataset are a relatively small sample, with 2,999 job transitions from 2012-2018. 
While these observations come from the high-quality HILDA dataset, which is Australia's pre-eminent and representative household survey~\cite{HILDA_2020}, it is nonetheless a relatively small sample to train a machine learning system against. 
A small sample enlarges the risks of biases emerging as the recommender system is dependent on a relatively small sample of observations to `learn' from and make predictions about future job transitions.
Longitudinal household surveys also suffer from panel attrition, including HILDA~\cite{watson2004sample}. 
However, this is only a minor issue for this study, as yearly job transitions are treated as independent observations. 
Our methods are not dependent on the longitudinal career pathways of individuals. 
With regards to the job ads data, we only had access to the Australian job ads dataset.
As a result, our analyses and results are specific to the Australian labor market.
However, this is a feature of our work rather than a limitation, as it allows to contextualize the analysis to geographical and temporal labor markets -- it has been shown that labor markets can be highly contextual~\cite{moro2021universal}.
One can easily leverage our methods to produce results for other countries by applying equivalent country-specific labor market data from job ads, employment statistics, and occupational transitions.

Second, the results presented in this paper have been \textit{aggregated to the occupational level}. 
That is, the explanatory features have been grouped by their 4-digit occupations, such as median salary and average education for a given occupation (see the \textit{Supplementary Information} for a full list of the features). 
Consequently, the job transition predictions in this paper are made at the grouped 4-digit occupational level for demonstration purposes, which does not differentiate within the same occupations. 
However, the flexibility of the methods presented in this paper can be applied at the individual level (or another arbitrary grouping), given the availability of appropriate data sources.

Third, there can be many factors that cause individuals to transition between jobs~\cite{MovingBetweenJobs}, beyond those used as explanatory variables in this study.
For example, it has been shown that personality profiles are predictive of different occupational classes~\cite{Kern2019-zg}.
Therefore, it is plausible that personality traits and values could influence not only the willingness to move between jobs but also the types of job transitions.
Similarly, there exists a Markovian assumption that a worker is described by the set of skills in their current job, therefore ignoring their past work experience and education.
Other factors such as competitive dynamics within specific industries and labor markets, macroeconomic conditions, and changes to individuals' household finances can all influence people transitioning between jobs.
Future work could look to incorporate these additional variables to help further explain and predict job transitions.

Last, we must acknowledge the risks of biases emerging from applying mechanical algorithms to `learn' from historical examples and make consequential recommendations to people, such as suggested career pathways. 
If the data used to train a machine learning system contains biases, then the predictions generated by the system are likely to reflect these biases. 
For example, there are structural biases in labor markets that influence employment outcomes, including biases based on gender~\cite{browne2003intersection}, race~\cite{pager2009discrimination}, age~\cite{carlsson2019age} and others. 
As the Australian labor market is not immune to systemic biases~\cite{foley2018does}, likely, the training data used for this research (HILDA -- a representative household survey) reflects these systemic biases to some extent. 
Therefore, the results presented in this paper should be viewed as `descriptive' of labor mobility in Australia rather than `prescriptive' of individuals' career options.
To help safeguard these systemic biases, we add a human-in-the-loop to filter the generated recommendations.
The system we design in \cref{fig:transition-example} is a decision-aid tool that filters top recommendations based on posting frequency (the link colors) to identify which occupations are growing in demand. 
Additional filters can be applied, such as top recommendations based on salary, education level, years of experience demanded, specific skill sets, industries, and others.
While these filters do not remove biases from the recommendations entirely, they do provide individuals using this system with greater autonomy in exploring potential career paths.

\section*{Conclusion}
Leveraging longitudinal datasets of real-time job ads and occupational transitions from a household survey, we have developed the \Method method to measure the distance between sets of skills. 
This enabled us to build systems that both \textit{recommend} job transition pathways based on personalized skill sets and \textit{detect} the growth of disruptive technologies in labor markets that could accelerate forced job transitions.
Our \textit{Job Transitions Recommender System} has the potential to assist workers, businesses and policy-makers to identify efficient transition pathways between occupations. 
These targeted and adaptive recommendations are particularly important during economic crises when labor displacement increases and workers are forced to transition to another job. 
The \textit{Job Transitions Recommender System} could therefore assist with the current labor crisis caused by COVID-19. Additionally, it could assist with potential future crises, such as accelerated job transitions caused by AI labor automation.

We further demonstrate the usefulness and flexibility of \Method by applying it as a measure of AI adoption in labor markets. 
This acts as an `early warning system' of forthcoming labor disruptions caused by the adoption and diffusion of AI within industries. 
Such a measure could complement other indicators of AI adoption, serving policy-makers and businesses to monitor the growth of AI technologies and its potential to accelerate job transitions.

While the future of work remains unclear, change is inevitable.
New technologies, economic crises, and other factors will continue to shift labor demands causing workers to move between jobs.
If labor transitions occur efficiently, significant productivity and equity benefits arise at all levels of the labor market~\cite{OECD2020}; if transitions are slow, or fail, significant costs are borne to both the State and the individual.
Therefore, it is in the interests of workers, firms, and governments that labor transitions are efficient and effective.
The methods and systems we put forward here could significantly improve the achievement of these goals.

\section*{Acknowledgments}
We thank Bledi Taska and Davor Miskulin from Burning Glass Technologies for generously providing the job advertisements data for this research and for their valuable feedback. 
We thank Stijn Broecke and other colleagues from the OECD for their ongoing input and guidance in the development of this work.
We thank Albert Carmon for his editing and feedback on language and style.
We acknowledge and thank Richard George from Faethm AI for facilitating access to the `non-essential' list of occupations in Australia during the initial stages of COVID-19.

\nolinenumbers

\clearpage

\clearpage
\appendix
\section*{Supporting information}
This document is accompanying the submission \PaperTitle by authors Nikolas Dawson, Mary-Anne Williams, and Marian-Andrei Rizoiu.
The information in this document complements the submission, and it is presented here for completeness reasons.
It is not required for understanding the main paper, nor for reproducing the results.

\subsection*{S1 Appendix: Related Work}
\label{related-work}
This section discusses the related works that have directly informed the research in \PaperTitle. There is firstly a discussion of the factors affecting labor mobility, the causes and effects of skill mismatches, measuring human capital transferability, and accounting for the asymmetries between jobs.
Then there is a brief literature review of the factors affecting the adoption of Artificial Intelligence (AI) technologies.

\subsubsection*{Job Transitions}
The related literature on job transitions broadly falls into the categories of labor mobility and human capital within the discipline of labor economics. `Labor mobility' refers to the allocation of workers to firms and their ability to move between jobs~\cite{borjas2010labor}. Labor mobility is an important determinant of healthy labor markets. Efficient labor movements enable firms to hire more productive workers, effectively match workers to jobs based on their preferences, and helps to protect markets against economic shocks and structural changes. `Human capital' refers to the skills, knowledge, capabilities, and experiences possessed by an individual that influence their productive capacities and that can be exchanged for labor at a prevailing market wage~\cite{Schultz1961-uq, becker1964human, Becker1990-zc}.

The process of labor mobility is constantly evolving and influenced by a variety of factors. These include labor market policies that shape hiring practices, job separation support programs, and relocation incentives~\cite{pries2005, bassanini2013, hassler2005}. It is also impacted by the extent of human capital in a labor market, which refers to the supply of skills, knowledge, and abilities of a labor force that firms can employ to produce goods and services~\cite{goldin2016}. According to Nedelkoska and Frank, skills should be considered part of human capital that is acquired through education, training, and work experiences~\cite{nedelkoska2019}. While access to education and training undoubtedly affects the acquisition of skills, particularly `general' skills~\cite{becker1964human}, skills are also acquired through work experience. Typically, firm or industry-specific skills are not perfectly mobile across employers and can hinder labor mobility~\cite{wasmer2006}.
Therefore, the extent of human capital `specificity' in labor markets is an important factor affecting labor mobility. It impacts how transferable skill sets are between jobs and reveals their underlying mismatch.
The remainder of this section reviews the literature relating to the cause, effects, and measurement of skill mismatch and skill transferability. 
There are subtle, but important, differences between these terms. `Skill mismatch' refers to the differences between the supply of and demand for skills in a labor market. Whereas `skill transferability' is the capacity to leverage previously acquired skills to perform tasks across different jobs, either because the tasks are similar or the skills can be flexibly applied to different tasks~\cite{nedelkoska2019}.The following literature forms the theoretical basis that directly informs our novel approach to measuring the distance between skills and sets of skills. 

\subsubsection*{Causes and effects of skill mismatches.}
Skills provide the means for workers to complete tasks that are required by jobs. A distinction should also be made between skills, knowledge areas, and abilities. `Skills' are the proficiencies developed through training and/or experience~\cite{Oecd2019-cl}; `knowledge' is the theoretical and/or practical understanding of an area; and `ability' is the competency to achieve a task~\cite{Gardiner2018-dt}, where a task is a unit of work required by a job. For simplicity, the term `skill' will collectively represent these three definitions throughout this paper.

Skill mismatch occurs when the skill demands of a job differ from the supply of skills available in a labor market~\cite{nedelkoska2019}. When labor demand outweighs the supply of specific skills, it is referred to as a `skill shortage'; when supply exceeds labor demand, it is often called `skill excess' or `over-supply'. Skill mismatches are closely monitored by labor economists due to the cost burdens they impose. For workers, they lower wages and employment opportunities; for firms, they restrict access to talent to implement specific tasks; for economies, they drain productivity~\cite{nedelkoska2019}.

The causes of skill mismatches can be both frictional and structural. There are search costs associated with a worker finding a job suitable to their skills, education, and experiences~\cite{topel1992}. The frictions of matching workers to appropriate jobs can hamper efficient job transitions and exacerbate skill mismatches. Structural factors causing skill mismatches, however, are concerned with the supply and demand for skills.

Regarding the supply side, the literature has mainly focused on the role of public institutions to facilitate skill development through education and training. Freeman, among others, demonstrated that the oversupply of skills can depress wage premiums that are typically earned by highly educated workers~\cite{freeman1975}. Further, Goldin and Katz showed that the wage premium of US college graduates was a function of the supply of college education, with college wage premiums increasing when the supply of college degrees in the labor market was low~\cite{goldin2009}.

Structural changes in the demand for skills are predominantly caused by (1) technological advances and (2) globalization or trade. Concerning technology and innovation, Vona and Consoli~\cite{vona2015innovation} present a useful framework for understanding the evolving relationship between skills and technological change. In the early stages of new technology adoption, the authors argue that tasks are typically complex and non-routine. Consequently, specialized and highly skilled labor is required to make productive use of these new technologies. As time progresses, however, knowledge becomes structured and codified, enabling tasks to be routinized and automated. Eventually, the marginal benefits of specialization diminish as the use of the technology becomes standardized and tasks are able to be performed by lower-skilled workers. Related, is the theory of \textbf{Skill-Biased Technological Change} (SBTC). The SBTC hypothesis posits that technologies disproportionately advantage highly-skilled labor over lower-skilled labor, as technologies tend to enhance the skills highly-skilled workers and automate lower-skilled workers~\cite{mincer1991human, berman1994changes, autor1998computing}. The SBTC hypothesis was later modified to account for the relationship between computers and task-specific requirements of jobs. This Task-biased Technological Change (TBTC) framework~\cite{acemoglu2011skills, autor2013putting} classifies labor tasks along two main spectra; routine to non-routine tasks and cognitive to manual tasks. Computerization, according to TBTC, tends to assist non-routine tasks and automate routine tasks. As a result, computers automate the labor tasks of middle-skilled workers (typically, routine-cognitive workers), which helps to account for recent dynamics such as declining real wages of middle-skilled workers and labor polarization~\cite{goos2014explaining}. However, the argument that the negative demand-side effects of computerization are limited to routine tasks is now coming under scrutiny. The rapid advances and diffusion of AI technologies cast doubt over this assumption. Brynjolfsson and McAfee~\cite{brynjolfsson2014second} and Frey and Osborne~\cite{Frey2017} present compelling arguments that the automation capabilities of AI are extending to non-routine tasks, both in the cognitive and manual domains. Non-routine tasks that were previously considered out of reach by AI are quickly outperforming human levels in a range of non-routine tasks, such as Natural Language Processing (NLP)~\cite{Brown2020-yu}, Image Recognition~\cite{Touvron2020-mz}, and unstructured learning tasks~\cite{Silver2018-kt}.

Globalization or trade also has important implications on the demand for skills, which can exacerbate skill mismatches. Offshoring enables firms to fulfill their required labor tasks without personal contact, which can be managed electronically without a loss in quality~\cite{blinder2013alternative}. This shifts the demand for skills towards countries with lower labor costs. Autor et al.~\cite{Autor2013-we} found that approximately one quarter of the decline in US manufacturing employment can be attributed to increasing trade with China.

Taken collectively, changes in the supply of and demand for skills alter the extent of skill mismatches in a labor market. This equilibrium is dynamic and directly affects labor mobility. The following subsection reviews the literature for measuring skill mismatch and transferability, which directly informs this work.

\subsubsection*{Measuring skill transferability and mismatch.}
This research is part of a small but growing area of labor economics that measures the `distance' between skills, jobs, and other defined skill sets. Among the earliest work in this area was conducted by Shaw~\cite{shaw1984formulation, shaw1987occupational} who defined measures of occupational distances via proxies of skill transferability across occupations. This was based under the assumption that occupations with high levels of skill transferability are strongly correlated with high probabilities of transitioning between these occupations. This is an assumption that we adapt, test, and prove  in \PaperTitle.

More recent studies have made use of skill and task-level data, such as the US Dictionary of Occupational Titles (DOT - a predecessor to O*NET) or the German Qualification and Career Survey (QCS). Poletaev and Robinson~\cite{poletaev2008human} use task-level data from DOT to study the similarity between occupations. The authors construct four measures of basic skills, applying the factor analysis method used by Ingram and Neumann~\cite{ingram2006returns}. These four skill measures characterize the `skill portfolios' of occupations, which are organized as vectors of skills. They then use Euclidean distance to compute the similarity between occupational skill vectors in order to identify which workers change their skill portfolios when transitioning between jobs. The authors show that workers who find jobs with similar skill requirements to their earlier jobs before displacement avoid large wage losses. 

Similarly, Gathmann and Schönberg~\cite{gathmann2010general} use the QCS to classify occupations into a 19-dimension skill space defined by the survey. Each occupation represents a skill vector, where occupations consist of certain skills with varying degrees of mastery. The authors use the angular distance between the 19 skill vectors to position the occupations and measure their relative distances. The authors demonstrate that individuals transition to occupations with similar task requirements and that the distance requirements decline with greater work experience.

Most recently, Alabdulkareem et al.~\cite{Alabdulkareem2018-jl} used techniques from Network Science and unsupervised Machine Learning to illustrate occupational polarization based on their underlying skill. Data sources included a combination of O*NET skill-level data and US occupational transitions data in the Current Population Survey from the US Bureau of Labor Statistics. The authors implemented an established measure from Trade Economics, called `Revealed Comparative Advantage' (RCA), to firstly measure the relative importance of a skill in a job while normalizing for high-occurring skills. After setting a threshold for skill importance, skill similarity was then calculated as the minimum of conditional probabilities that a skill pair are both important in a job when they co-occur. The authors then used these pairwise skill similarities to map workplace skills as a network, highlighting skill polarization and proving a correlation with wage polarization. Dawson et al.~\cite{dawson2019adaptively} extended this approach by applying this method to real-time job ads data to adaptively select occupations based on their underlying skill demands. This enabled the authors to accurately monitor changing labor demands and detect skill shortages for an evolving set of Data Science and Analytics occupations in Australia. The skill similarity methods applied by Alabdulkareem et al.~\cite{Alabdulkareem2018-jl} and Dawson et al.~\cite{dawson2019adaptively} provide the foundation for the \Method method.

While all of these approaches represent significant contributions in the evolution of measuring skill transferability, there is one major shortcoming. All of these methods yield symmetric distance measures. That is, the distance from one skill or occupation to another is the same despite the direction. For example, according to these methods, it is just as difficult for a Nurse to become a Surgeon as the other way around. Intuitively, however, acquiring certain skills to transition to a particular occupation is more difficult in one direction than the other. In this sense, skill acquisition and occupational transitions are directed and asymmetrical.

\subsubsection*{Asymmetric measures for skill mismatches.}
Nedelkoska et al.~\cite{nedelkoska2015skill} develop skill mismatch metrics that account for the strong asymmetries in the transferability between skills. The authors construct occupational skill profiles by using factor analysis to extract five task-based skills on German administrative and data on individuals' work histories. They then calculate the share of workers in each occupation carrying out these tasks. The average years of education and training associated with each task are used as weights to indicate skill complexity required by an occupation. Adding these weights reveals asymmetries between skills and therefore occupations. They show that by switching occupations, people incur both skill shortages and skill redundancies, which results in significant wages losses up to 15 years following the job displacement. While accounting for skill asymmetries represents a clear improvement for measuring the distance between skills, using years of education and training as the sole proxy for skill complexity is questionable. As previously stated, work experiences are an important contributor to the acquisition of skills and causes of mismatch.

Bechichi et al.~\cite{bechichi2018} adapt the Nedelkoska et al.~\cite{nedelkoska2015skill} model by analyzing occupational data from the OECD Survey of Adult Skills (PIAAC). They firstly use the six task-based skill indicators from PIACC~\cite{grundke2017having}. The authors then measure these indicators on 127 occupations (at the 3-digit occupational level) across 31 different countries to assess occupational distances based on `cognitive skills' and skills acquired from tasks `on the job'. This method accounts for skill asymmetries and skills acquired from work experiences. The resulting `skill shortage' and `skill excess' measures are then used to predict education and training resources required to transition workers from one occupation to another. Therefore, this research represents another advance toward the goal of accurately measuring skill and occupational distances. However, a minor shortcoming of this work is that it is performed at the 3-digit occupational level, which is a relatively high classification level (1-digit being the highest and 6-digit being the lowest and most detailed). Additionally, surveys provide lagging data that are typically slow to report and expensive to conduct. This is problematic in labor crises, such as the job displacements caused by COVID-19. Dynamics of labor markets quickly change in times of crisis and displaced workers are faced with transitioning between jobs with rapidly evolving skill demands. Real-time data, therefore, becomes essential.

Our research builds on these significant works and addresses both of these shortcomings by using real-time job ads data and applying a method capable of measuring the distance between any defined set of skills, such as occupations at the detailed 6-digit occupational level, industries, or even personalized skill sets.

\subsection*{S2 Appendix: Artificial Intelligence Adoption}
The labor market impacts of AI depend on the adoption rates of AI technologies by firms.
If firms are slow or fail to adopt AI, then its effects are naturally restricted. 
Therefore, the risks of AI accelerated labor automation will only be realized if these technologies are adopted by firms, absorbed in workflows, and broadly diffused. Otherwise, they're just isolated use cases.

This consideration, however, is often ignored. Much of the recent research on the economic impacts of AI assume broad adoption and diffusion. 
For example, the prominent study by Frey and Osborne estimated that 47\% of occupations face a near-term risk of automation from AI~\cite{Frey2017}. These results were based on the assessments of a small panel of Machine Learning experts who were asked to identify which of 70 jobs were `completely automatable' in 2013. 
However, these forecasts rely on some questionable assumptions.
Chief among them is that firms will quickly and efficiently adopt AI for commercial use. This should not be taken as a given. 
As Bessen et al.~\cite{Bessen2002, Bessen2018} point out that, just because new technologies have commercial applications does not mean that they will be adopted and diffused in a timely manner. Therefore, understanding the factors that influence the adoption and diffusion of AI in firms is important. It enables more accurate forecasting and better planning for policymakers, businesses, and civil society.

\subsubsection*{Explanatory variables for AI adoption and diffusion.} 
Research on the factors that affect firms' decisions to adopt digital technologies is well established~\cite{rogers1976new, karahanna1999information, im2011international}. Researchers have closely examined the adoption dynamics of innovations such as personal computers~\cite{thong1999integrated}, the Internet~\cite{andres2010diffusion}, and social media~\cite{perrin2015social}. AI builds upon these digital technologies. The factors that influence the adoption of AI by firms differ by degree but not by kind. The literature suggests eight major factors influencing AI adoption rates at the firm-level:

(1) \textit{Competition}: McKinsey Global Institute found that the extent of rivalry within markets has the largest effect on AI adoption~\cite{Bughin2018-qr}. This is consistent with game theory~\cite{moorthy1985using}, where the marginal propensity to adopt AI depends on the proportion of rivals that have already decided to adopt. Assuming the new technology becomes broadly diffused, then early adopters typically enjoy disproportionate rewards. However, as more firms adopt, the marginal incentive to adopt diminishes as the technology provides less competitive advantages. Therefore, laggard firms are punished with shrinking market shares~\cite{Andrews2015-yy}. These competitive forces drive adoption rates as firms jostle to assert a competitive edge and advance market share~\cite{Mamer1987-cn}. However, adoption decisions are made with imperfect information as it can be difficult to know what actions competitors are taking. Competition, therefore, can drive rapid periods of adoption growth. 

(2) \textit{Firm characteristics}: The size, income level, and industry of firms have all been shown to affect the rate that a new technology is adopted~\cite{hall2003adoption}. For example, larger firms, by headcount and income, typically adopt digital technologies earlier and at faster rates than smaller firms. Also, firms in Financial Services and ICT industries tend to adopt digital technologies earlier and at faster rates than firms in Agricultural and Construction industries~\cite{ABStech}. Similarly, the AI adoption indicator we propose suggests material differences between industry categories, with highest levels of adoption in Financial \& Insurance Services firms and lowest in the Agriculture Industry.

(3) \textit{Labor force skill capabilities}: Emerging technologies, such as AI, often require specific skills~\cite{Bessen2002}. The availability of workers with these skills can influence the extent of adoption and diffusion~\cite{beaudry2006endogenous}. The ability to access such labor competencies, however, varies between firms, industries, and economies. The implementation of AI requires strong technical competencies. These competencies are unevenly distributed between firms, industries, and economies~\cite{Bessen2018}. Therefore, the more firms are able to access relevant skilled labor, the greater the likelihood that firms will adopt AI.

(4) \textit{Digital Maturity}: Previous research has shown that the adoption of new digital technologies often depends on the adoption of previous digital technologies~\cite{Andrews2018-wt}. For instance, broadband infrastructure supports the adoption of more sophisticated digital applications. This relationship also appears to hold for AI. According to McKinsey Global Institute, firms that have adopted and absorbed cloud infrastructure and `web 2.0 technologies', such as mobile technologies and Customer Relationship Management (CRM) systems, are more likely to adopt AI technologies~\cite{Bughin2018-qr}.

(5) \textit{Expected return on AI investments}: Firms' perceptions of the value that a new technology can create also influences adoption rates~\cite{anderson2004information}.
Similarly, firms that are positive about the business use cases of AI are more likely to adopt earlier and faster~\cite{Bughin2018-qr}. 
Conversely, firms that are uncertain about AI use cases are slower or less likely to adopt, which delays aggregate adoption rates.

(6) \textit{AI complements}: As with other General Purpose Technologies, the more complementary technologies are developed and implemented, the faster AI will be adopted by firms~\cite{Brynjolfsson2018-rs}. That is, the more a firm invests in one type of AI, the more likely it will invest in another. For example, a retailer that implements robotic process automation to retrieve stock is more likely to adopt computer vision to identify inventory items than a retailer that has not adopted any AI technologies. Capital investment deepens as AI is increasingly absorbed in workflows.

(7) \textit{Regulatory effects}: Regulatory effects can be important to consider when comparing the adoption rates across economies~\cite{im2011international, perino2012does}. For example, it is plausible that the more stringent data protection regulations in Europe could delay AI adoption in European firms compared to US firms in the short-run.

(8) \textit{Standardization and usability}: As the use of emerging technologies are standardized across firms and industries, the ease of use for these technologies naturally improves, which has been shown to accelerate adoption and diffusion rates~\cite{Bessen2015-pp}. While AI models are still `narrow' in the sense that they tend to be highly specific to a particular task and require non-routine customization (as in hyper-parameter tuning or feature data engineering), AI usability has improved over the past decade. For example, individuals are able to implement high-performing machine learning models using their own data with little or no knowledge of scripting languages (see \cite{GoogleCloud}). As the use of AI technologies become standardized and usability improves, it is likely that this will increase adoption rates.

While other variables could affect rates of AI adoption, the eight factors listed above are likely to account for a significant proportion of firm-level AI adoption decisions.

\subsection*{S3 Appendix: Using a standardized occupation taxonomy -- ANZSCO.}
All data sources mentioned above correspond to their respective occupational classes according to the Australian and New Zealand Standard Classification of Occupations (ANZSCO)~\cite{ANZSCO2013}.
ANZSCO provides a basis for the standardized collection, analysis and dissemination of occupational data for Australia and New Zealand. 
The structure of ANZSCO has five hierarchical levels - major group(1-digit), sub-major  group (2-digit), minor group (3-digit), unit group (4-digit) and occupation group (6-digit). 
For visualizing the distance between occupations in Fig 1.B in \PaperTitle, the 6-digit grouping level was applied. We used the 6-digit level here because (1) it is the most detailed occupational grouping and (2) the automation probabilities from Frey \& Osborne were mapped at this level~\cite{Frey2017}.
The presented results for the `Job Transitions Recommender System' and the subset of the \textit{Transitions Map} shown in Fig. 3 are grouped at the 4-digit unit grouping level. Occupational recommendations were made at this level to match the ground-truth of actual transitions from the Household, Income and Labour Dynamics in Australia (HILDA) longitudinal dataset~\cite{HILDA_2020}. 

\subsubsection*{Shortcomings of ANZSCO.} There are shortcomings to analyzing occupations within ANZSCO classifications. 
Official occupational classifications, like ANZSCO, are often static taxonomies and are rarely updated. 
They therefore fail to capture and adapt to emerging skills, which can misrepresent the true labor dynamics of particular jobs. 
For example, a `Data Scientist' is a relatively new occupation that has not yet received its own ANZSCO classification. 
Instead, it is classified as an `ICT Business \& Systems Analyst' by ANZSCO, grouped with other job titles like `Data Analysts', `Data Engineers', and `IT Business Analysts'. 
However, as ANZSCO is the official and prevailing occupational classification system, all data used for this research are in accordance with the ANZSCO standards.

\subsubsection*{Mapping O*NET to ANZSCO.}
To leverage the strengths of earlier research by Frey and Osborne~\cite{Frey2017} on the occupational risks of labor automation caused by AI technologies, we first needed to map O*NET occupations to ANZSCO, so as to take advantage of their automation risk probabilities at the 6-digit level. 
O*NET is a standardized and publicly available database of labor market data in the United States~\cite{US_Department_of_Labor_undated-oe}. The occupations, however, are slightly different compared to ANZSCO.
Therefore, we used a concordance table from the Australian Federal Department of Education, Skills and Employment~\cite{ONET_ANZSCO_concordance} to map O*NET occupations to ANZSCO at the 6-digit level. This resulted in each ANZSCO occupation being assigned an automation risk probability according to the Frey and Osborne research.

\subsection*{S4 Appendix: Model Features}
\label{subsec:model-features}

Below is a summary of the features included within the `Job Transitions Recommender System' models.
The features have been grouped into `Labor Demand' (job ads data) and `Labor Supply' (employment statistics) categories for ease of review.
Each feature in \nameref{tab:feature-summary} is measured at the ANZSCO 4-digit level per calendar year from 2012-2018.
This reflects the first available year of the longitudinal job ads data (2012) and the most recent year of the `ground-truth' HILDA data (2018).
The `source' and `target' occupations are independently associated with each of the features in \nameref{tab:feature-summary}. 
However, the `theta' (distance between skill sets measure) and `Difference' features relate to both the `source' and `target' occupational pair. In total there are 19 features.

\paragraph*{S1 Table} \label{tab:feature-summary}
\textbf{Summary of constructed features and their explanation.}
\\[12pt]
{	\centering
	\small
    \begin{tabular}{crp{6.3cm}}
        \toprule
        & Feature & Description \\ \midrule
        \multirow{9}{*}{\rotatebox[origin=c]{90}{Labor Demand}} 
        & theta: & \Method{} distance between `source' and `target' occupation \\
        & Posting Frequency: & number of job advertisement vacancies \\
        & Posting Frequency Difference: & difference between the `source' and `target' posting frequency \\
        & Median Salary: & maximum median salary advertised \\
        & Salary Difference: & difference between the `source' and `target' salaries \\
        & Minimum Education: & minimum years of formal education required \\
        & Education Difference: & difference between the `source' and `target' years of formal education required \\
        & Minimum Experience: & minimum years of formal experience required \\
        & Experience Difference: & difference between the `source' and `target' years of experience required \\
       \midrule
       \multirow{4}{*}{\rotatebox[origin=c]{90}{Labour Supply}} 
        & Total Employed: & total number employed at ANZSCO Unit level (000's)\\
        & Total Employed Difference: & difference between the `source' and `target' of total number employed at ANZSCO Unit level (000's)\\
        & Total Hours Worked: & total hours worked at ANZSCO Unit level (000's)\\
        & Total Hours Worked Difference: & difference between the `source' and `target' of total hours worked at ANZSCO Unit level (000's)\\
        \bottomrule
    \end{tabular}
}

\newpage
\subsection*{S5 Appendix: Validation}
\subsubsection*{Statistical Test.}

\begin{figure}
    \centering
    \includegraphics[width=0.9\textwidth]{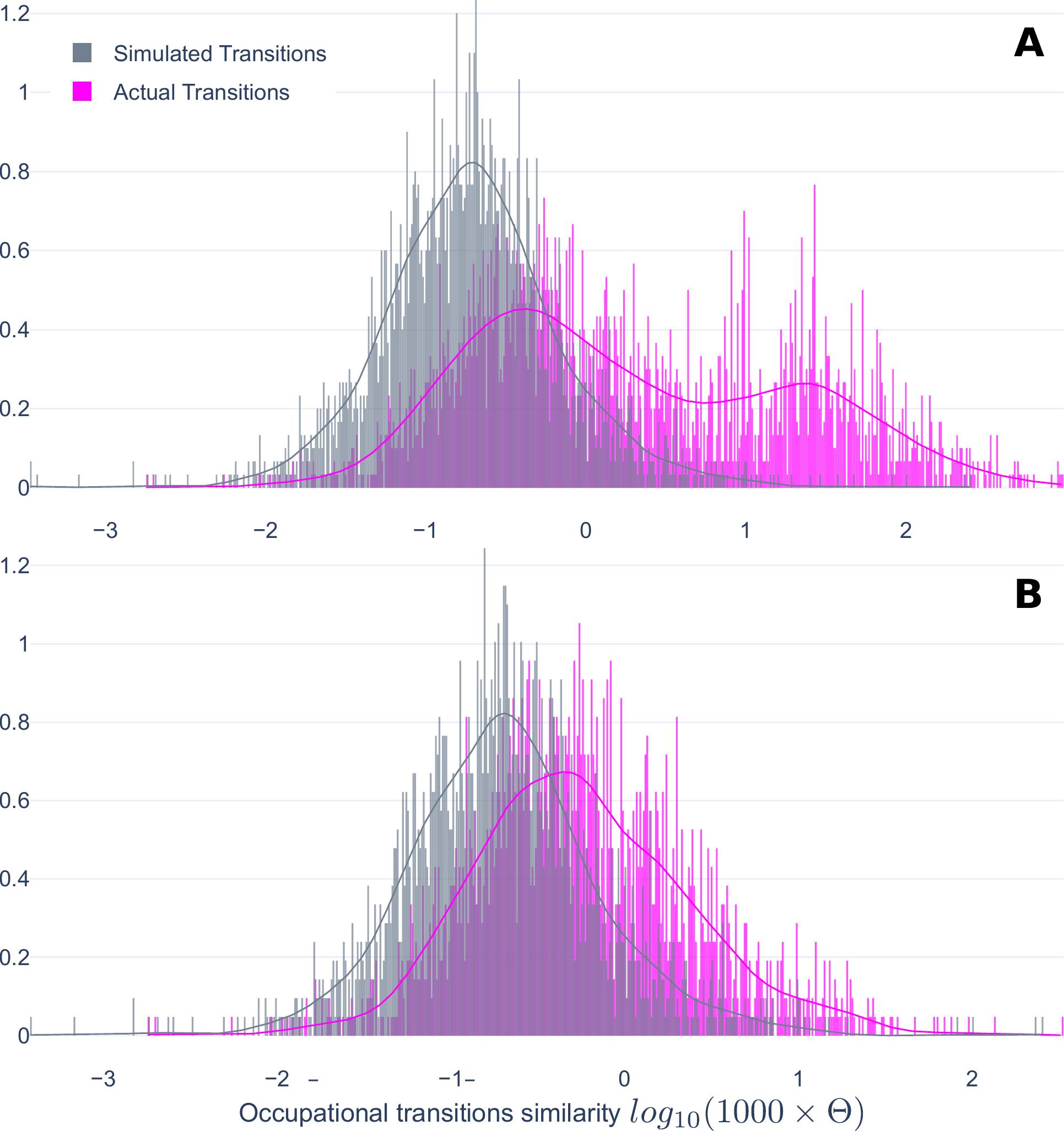}
    \caption{Density plots for the statistical tests against all occupational transitions (A) and against transitions where the worker changed occupations (B).}
    \label{S1_Fig}
\end{figure}

To obtain initial validation of the \Method distance measures, we conducted a paired statistical test, as explained in \PaperTitle. 
To run this experiment, we labeled each `source' and `target' occupational pair with their distance measure for their given year (called the `True Sample').
We then simulated an alternate sample of transitions where we maintain the same `source' occupations and randomly select `target' occupations, all assigned with their pairwise distance scores (called `Simulated Sample').
\cref{S1_Fig}-A shows the distribution of all job transitions, including transitions to the same occupation. We found that the differences between the `True' and `Simulated' transition samples are statistically significant (t-statistic = $16.272$, p-value = $2.707 \times 10^{-58}$, Cohen's D effect size = $0.42$). 
However, 909 of the 2999 (or 30\%) of the yearly job transitions from 2012-2018 are movements to the same occupation. Intuitively, the skill set distance of transitioning to another job in the same occupation is likely small, especially compared to other occupations.
Therefore, we wanted to test whether the statistical significance holds when we exclude transitions to the same occupation. 
To run this test, we first removed job transitions to the same occupation (leaving 2090 occupations from 2012-2018). 
Following the same process described above, we created `True' and `Simulated' samples.
\cref{S1_Fig}-B illustrates the differences between the two samples, which are again statistically significant (t-statistic = $4.514$, p-value = $6.535 \times 10^{-06}$); however, the effect size is lowered (Cohen's D effect size = $0.14$).

We repeated the procedure 100 times: we generated 100 `Random' populations and we perform the statistical testing for each. 
As shown in \PaperTitle, 87 of the 100 obtained p-values were lower than 0.05 providing confidence in the statistical significance of \Method to represent occupational transitions in this research.

\newpage
\subsubsection*{Job Transitions Recommender System validation.}

\begin{figure}
    \centering
    \includegraphics[width=0.9\textwidth]{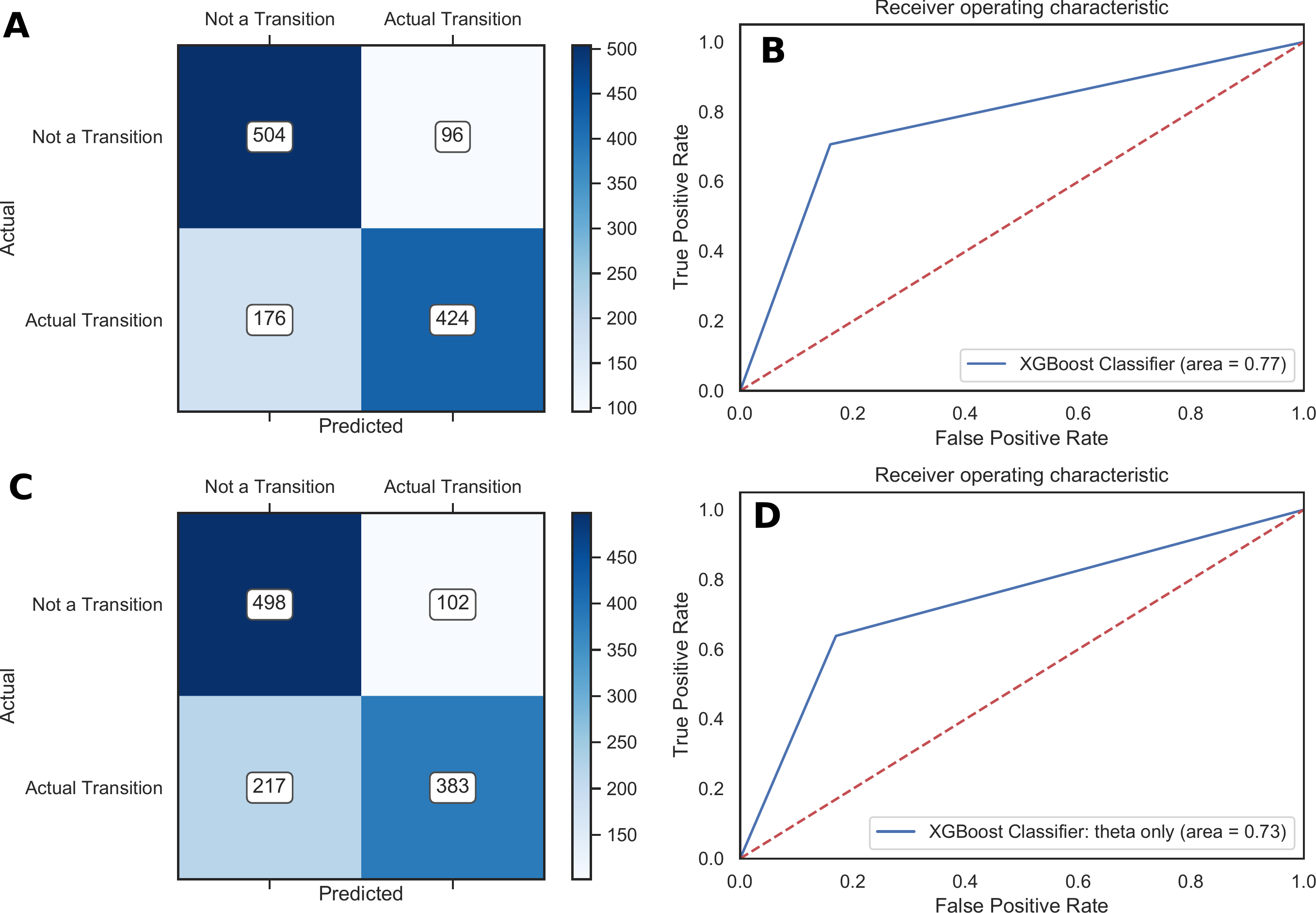}
    \caption{\textbf{Prediction performance and confusion matrix.} Job transition model that includes all features achieves the highest results, as seen with (A) the confusion matrix and (B) the ROC curve; whereas the job transitions classifier model that only includes \Method distance method has lower performance, as seen by (C) the confusion matrix and (D) the ROC curve.}
    \label{S2_Fig}
\end{figure}

\cref{S2_Fig}-A shows the confusion matrix for the binary classifier model containing all of the features from \nameref{tab:feature-summary}. This feature configuration achieved the highest performance (Accuracy = 76\% and F1 Macro Average = 77\%). As observed, this trained model was able to predict \textit{True Negatives} (`Not a Transition' -- Recall = 84\%) slightly better than \textit{True Positives} (`Actual Transition' -- Recall = 71\%).
\cref{S2_Fig}-B shows the `receiver operating characteristic' curve (ROC curve), which is the performance of the binary classification model at all classification thresholds. ROC curves summarize the trade-off between the \textit{True Positive} rate (y-axis) and \textit{False Positive} rate (x-axis) for the classifier model using different probability thresholds. Generally, high-performing models are represented by ROC curves that bow up to the top left of the plot. 
As illustrated in \cref{S2_Fig}-B, the blue ROC curve is consistently above the diagonal red dashed line that represents a 50\% probability -- models that perform below this dashed line are no better than random chance. This reinforces that the model consisting of `All Features' achieves strong performance levels.

Similarly, \cref{S2_Fig}-C shows the confusion matrix and \cref{S2_Fig}-D illustrates the ROC curve for the classifier model that only includes the \Method distance measure (`theta'). While the `theta only' model still performs relatively well (Accuracy \& F1 Macro Average = 73\%), performance does decline. Again, the \textit{True Negatives} (`Not a Transition' -- Recall = 83\%) outperform the \textit{True Positives} (`Actual Transition' -- Recall = 64\%). This highlights that the added labor market features from job ads data and employment statistics increased the performance capabilities of the models to predict \textit{True Positives}. These performance differences are also represented in \cref{S2_Fig}-D showing a slightly lower ROC curve for models that used `theta' alone.

\newpage
\subsubsection*{Ablation Test and Feature Importance.}

\begin{figure}
    \centering
    \includegraphics[width=0.9\textwidth]{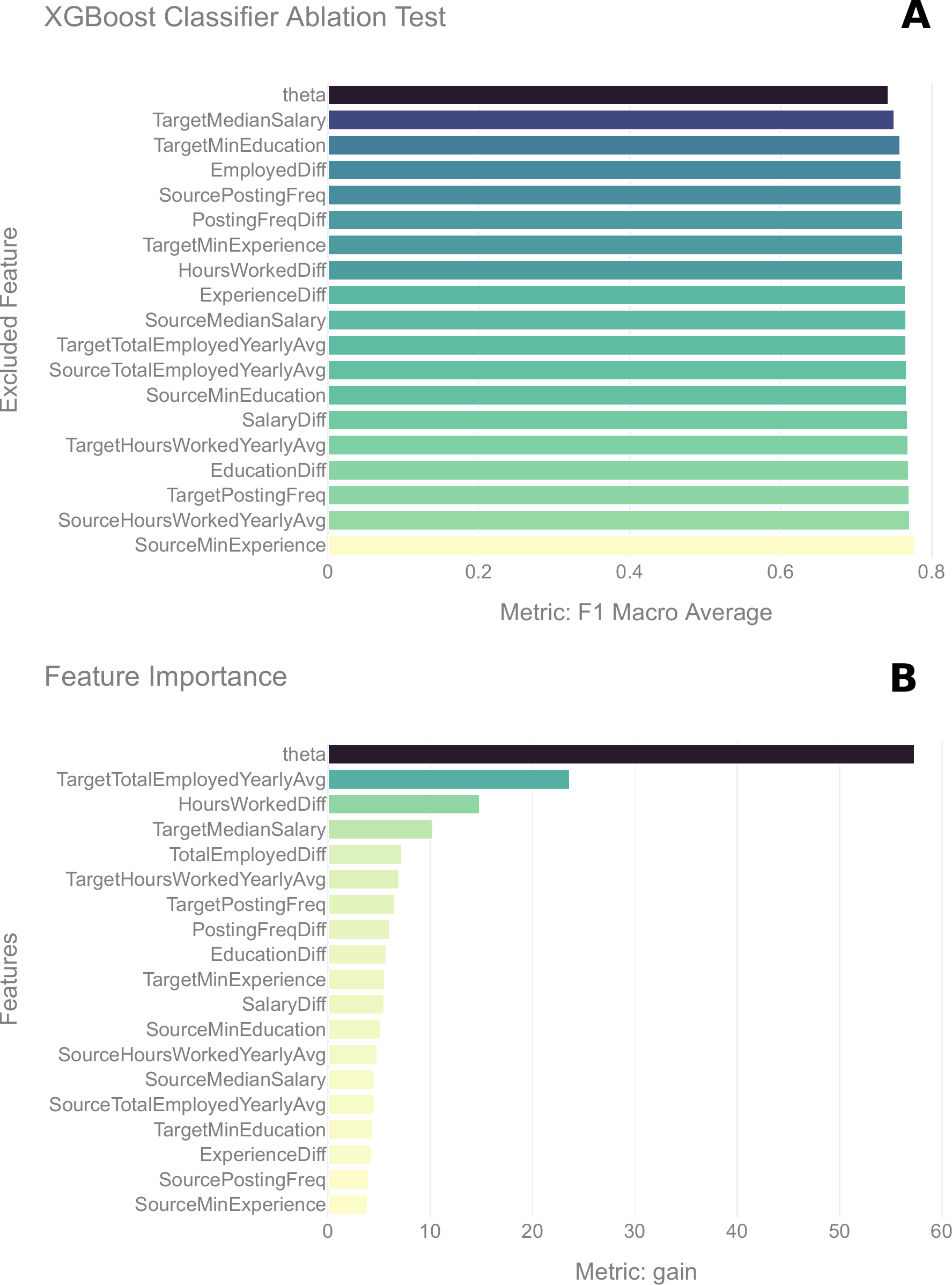}
    \caption{\textbf{Quantify feature importance.} 
(A) Ablation test of classifier features and (B) feature importance analysis both show that the \Method distance measure (`theta') is the most important feature for predicting occupational transitions.}
    \label{S3_Fig}
\end{figure}

In order to understand the relative importance of the modeled features in the `Job Transitions Recommender System', we conducted an ablation test and feature importance analysis. An ablation test involves iteratively removing one feature from the feature set and then re-training the model to make predictions and evaluate performance. We conclude that a feature is `more important' to a model's predictive capabilities if performance declines when it is removed. 
\cref{S3_Fig}-A shows the results of all 19 features, highlighting that the largest performance decline occurred when the `theta' distance measure was removed. These models were all trained with a consistent setup, as explained in \PaperTitle.

To reinforce the results from the ablation test, we then conducted a feature importance analysis as seen below.
We use the `Gain' metric, which shows the relative contribution of each feature to the model by calculating the features' contribution for each tree in the XGBoost model. A higher gain score indicates that a feature is more important for generating a prediction. 
Again, `theta' is overwhelmingly identified as the most important feature for predicting job transitions.

\newpage
\subsection*{S6 Appendix: Recommending Jobs \& Skills}

\paragraph*{S2 Table} \label{S2_Table} 
\textbf{Transitions Example -- Domestic Cleaner}
\\[12pt]
{\centering
\begin{adjustbox}{max width=1\textwidth,center}
\begin{tabular}{lrrrrr}
\toprule
                  Occupation &  Transition Probability &  Num. Job Ads 2019 &  Num. Job Ads 2020 &  Difference &  Percentage Difference \\
\midrule
           Domestic Cleaners &       0.960395 &         323 &         276 &   -47 & -14.551084 \\
         Commercial Cleaners &       0.946621 &         865 &         671 &  -194 & -22.427746 \\
                     Waiters &       0.943874 &         690 &         253 &  -437 & -63.333333 \\
 Bar Attendants and Baristas &       0.937961 &         600 &         180 &  -420 & -70.000000 \\
  Sales Assistants (General) &       0.935315 &        2835 &        1609 & -1226 & -43.245150 \\
                       Chefs &       0.926472 &        1904 &         877 & -1027 & -53.939076 \\
                       Cooks &       0.914349 &         726 &         356 &  -370 & -50.964187 \\
    Aged and Disabled Carers &       0.893725 &         961 &        1302 &   341 &  35.483871 \\
                Child Carers &       0.887601 &         837 &         414 &  -423 & -50.537634 \\
              General Clerks &       0.876921 &        2281 &        1466 &  -815 & -35.729943 \\
\bottomrule
\end{tabular}
\end{adjustbox}
}

\subsection*{S7 Appendix: AI Adoption}

\paragraph*{S3 Table} \label{S3_Table} 
\textbf{AI Similarity Scores.}
The table below contains the underlying data for the AI Adoption radar chart in the \textit{Developing a Leading Indicator of AI Adoption} section.
\\[12pt]
{\centering
\begin{adjustbox}{max width=1\textwidth,center}
\begin{tabular}{lrrrr}
\toprule
                                        Industry &     2013 &     2016 &     2019 &  Percentage change 13-19 \\
\midrule
                Financial and Insurance Services &  0.000958 &  0.001599 &  0.002887 &            201.395294 \\
        Information Media and Telecommunications &  0.001057 &  0.001283 &  0.002286 &            116.285278 \\
 Professional, Scientific and Technical Services &  0.000545 &  0.001027 &  0.001590 &            191.537693 \\
                                    Retail Trade &  0.000266 &  0.000568 &  0.001348 &            407.375732 \\
      Electricity, Gas, Water and Waste Services &  0.000443 &  0.000520 &  0.001300 &            193.594618 \\
                          Education and Training &  0.000707 &  0.000933 &  0.001257 &             77.744881 \\
               Transport, Postal and Warehousing &  0.000282 &  0.000427 &  0.000981 &            247.957226 \\
                                    Public Admin &  0.000270 &  0.000481 &  0.000905 &            234.594611 \\
         Rental, Hiring and Real Estate Services &  0.000157 &  0.000439 &  0.000775 &            392.117595 \\
                    Arts and Recreation Services &  0.000256 &  0.000571 &  0.000738 &            188.519596 \\
                                   Manufacturing &  0.000206 &  0.000356 &  0.000690 &            235.589203 \\
             Administrative and Support Services &  0.000243 &  0.000346 &  0.000661 &            172.344264 \\
                                 Wholesale Trade &  0.000219 &  0.000388 &  0.000642 &            193.307221 \\
                                          Mining &  0.000260 &  0.000223 &  0.000595 &            128.900314 \\
                                    Construction &  0.000136 &  0.000236 &  0.000487 &            257.990332 \\
                                  Other Services &  0.000136 &  0.000180 &  0.000306 &            124.906123 \\
               Health Care and Social Assistance &  0.000080 &  0.000170 &  0.000281 &            252.858476 \\
                 Accommodation and Food Services &  0.000086 &  0.000208 &  0.000267 &            208.987030 \\
               Agriculture, Forestry and Fishing &  0.000108 &  0.000183 &  0.000239 &            120.979250 \\
\bottomrule
\end{tabular}
\end{adjustbox}
}

\FloatBarrier

\noindent\textbf{Temporal AI skill similarity to Australian Industries.}
The figure below is another visualization of the same data in the \textit{Developing a Leading Indicator of AI Adoption} section illustrating all of the yearly AI similarity scores for the Industries from 2012-2019.

\begin{figure*}
    \centering
\includegraphics[width=1\textwidth]{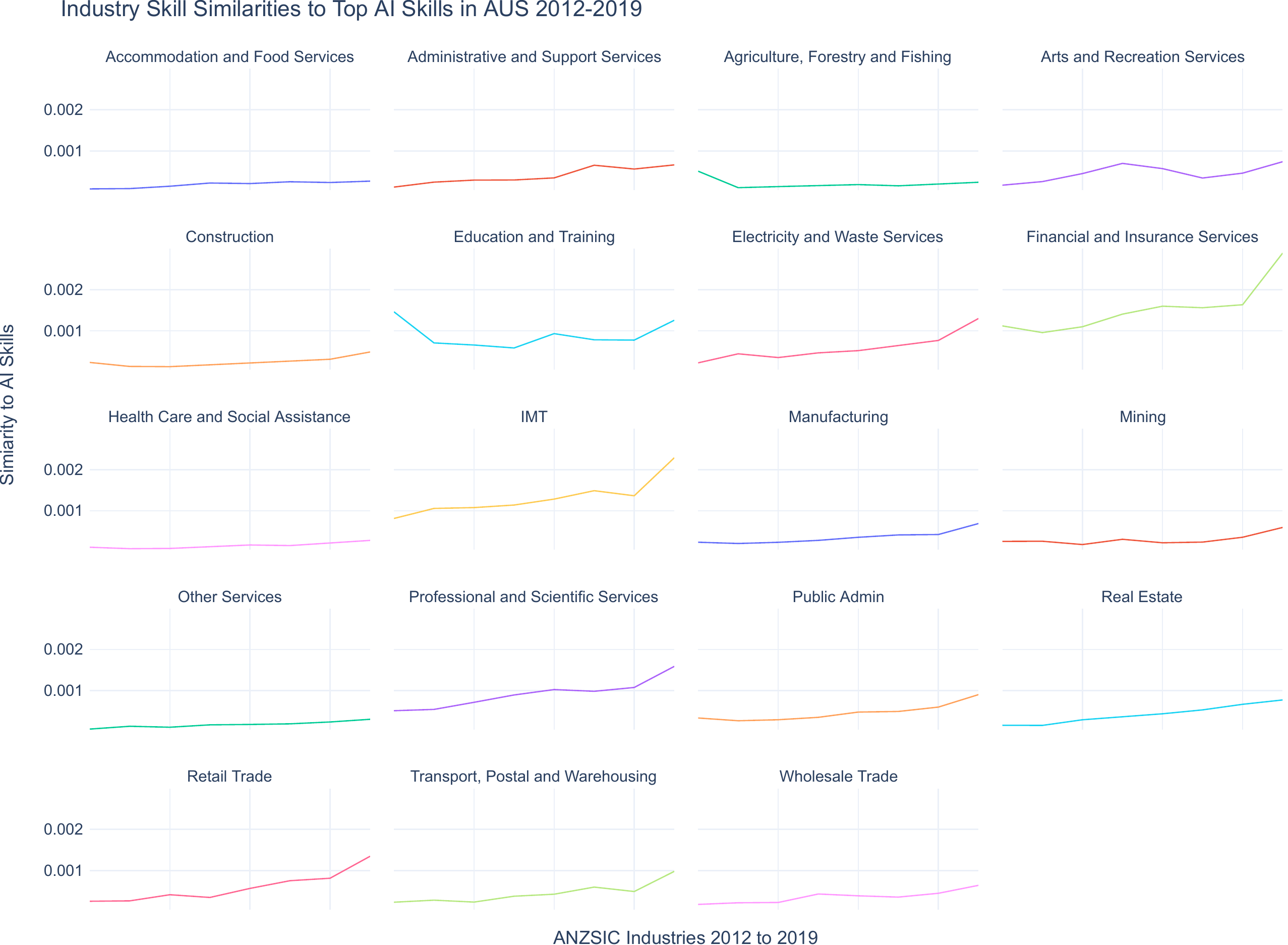}
\caption{Yearly skill similarities between AI skills and Australian Industry (ANZSIC Division) skill sets from 2012-2019.}
    \label{S4_Fig}
\end{figure*}

\FloatBarrier

\subsection*{S8 Appendix: Skill Count Distribution}
As discussed in \textit{Materials \& Methods}, before calculating individual skill similarities, we filtered out extremely rare skills to reduce noise and computational complexity. We set the minimum yearly skill count threshold to be greater than or equal to 5.

\begin{figure*}
    \centering
\includegraphics[width=0.9\textwidth]{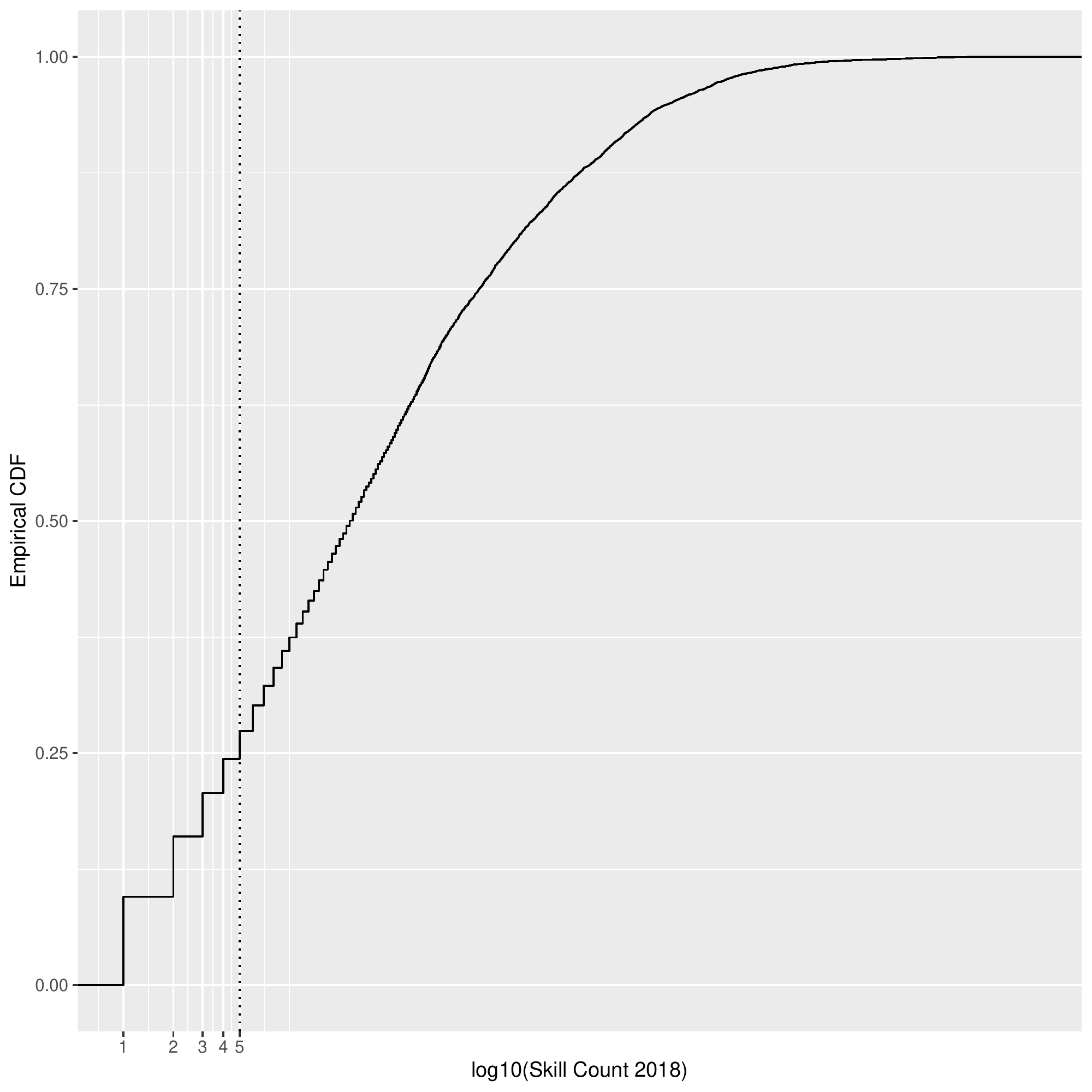}
\caption{Empirical Cumulative Distribution Function of skill counts within job ads for 2018.}
    \label{fig:ecdf}
\end{figure*}

As seen in the Empirical Cumulative Distribution Function (ECDF) in \cref{fig:ecdf}, this threshold represents over 75\% of all skills in 2018 (6,981 skills). All skills on the left side of the dotted threshold line were excluded, which accounted for $<$25\% of skills.

\FloatBarrier

\subsection*{S9 Appendix: Posting Frequency of AI Seed Skills}
\begin{figure*}
    \centering
        \centering
        \includegraphics[width=0.9\textwidth]{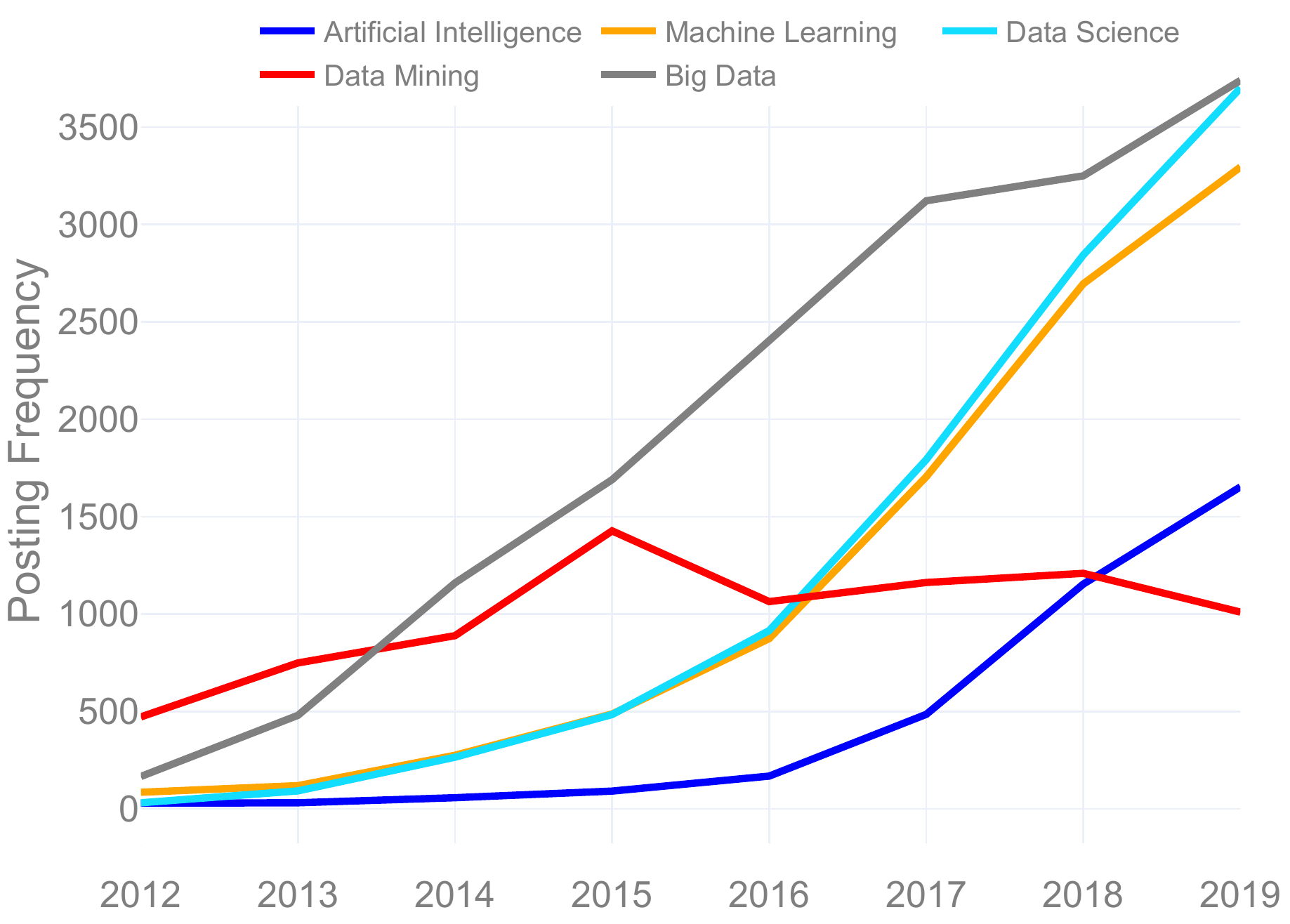}
    \caption{\textbf{Posting frequency of AI skills.}
Yearly posting frequency of the five AI seed skills used to build a dynamic list of yearly AI skills.}
    \label{fig:ai-seed-freq}
\end{figure*}

We selected five `seed skills' to construct a dynamic list of yearly AI skills, as described in Sec. \textit{Materials \& Methods} and visualized in Fig. 6 in the main paper. 
This allowed us to measure the distance between AI skills and industry skill sets, capturing both the evolution of skill demands and accounting for skill importance. 
The most common method, however, is to simply count the frequency of a skill (or group of skills) over time. 
In \cref{fig:ai-seed-freq}, we show the posting frequency of the five AI seed skills used to construct the dynamic list of yearly AI skills. The five AI seed skills being: (1) Artificial Intelligence; (2) Machine Learning; (3) Data Science; (4) Data Mining; and (5) Big Data.

As \cref{fig:ai-seed-freq} shows, the posting frequency for all five seed skills increases from 2012 to 2019, albeit at different rates. 
`Data Science' experiences the steepest increases over this period. 
Whereas `Data Mining' has had more modest growth, reaching its highest posting frequency levels in 2015 and has since declined.

\begin{figure*}
    \centering
        \centering
        \includegraphics[width=0.9\textwidth]{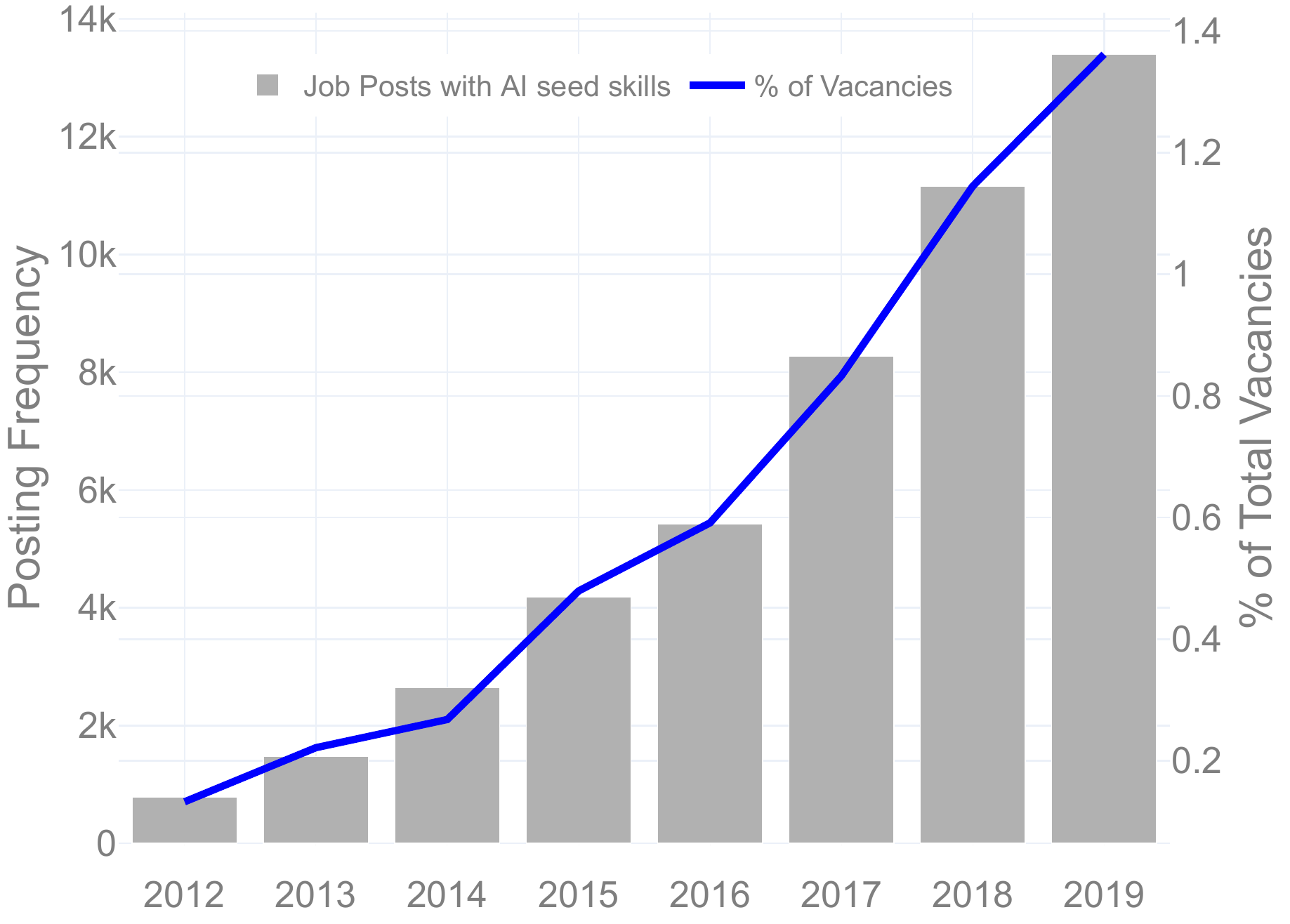}
    \caption{\textbf{Vacancy rate of AI skills.} 
The percentage of vacancies in Australia that contain these five AI seed skills.}
    \label{fig:ai-seed-perc-vacancies}
\end{figure*}

\cref{fig:ai-seed-perc-vacancies} shows that not only have the absolute posting frequencies of the AI seed skills increased, but also the percentage of vacancies containing these skills. 
In 2012, approximately 0.13\% of Australian job ads (or 783 vacancies) contained at least one of the AI seed skills. 
In 2019, this had increased to more than 1.3\% of job ads (or 13,399 vacancies) -- growth of over ten times the percentage of job ads requiring the AI seed skills. 

While these simple metrics provide an indication of the degree of growth required by the AI seed skills, there are some fundamental shortcomings for using posting frequencies as a proxy for labor demand. 
A discussion on the disadvantages of using posting frequency compared to skill similarity follows.

\newpage
\subsection*{S10 Appendix: Advantages of Skill Similarity over Posting Frequency}
The proxy most widely used in literature~\cite{Carnevale2014-xc} for skill importance is skill frequency. This simply counts how many times a skill appears in job ads associated with a given occupation (or other groups) during a predetermined period of time; the higher the count, the greater the demand and, implicitly, the greater importance of the skill to the occupation.
While skill frequency can provide some indication of labor demand, it fails to normalize for skills that are demanded by all or most jobs. 
This does not necessarily reveal which skills are more or less important to a given occupation, as some skills generalize across all occupations at high frequencies. For example, `Communication Skills' and `Teamwork' occur in over one-quarter of all job ads and are ubiquitous across all occupations).
However, we know that some skills are more important than others to specific jobs.
We therefore capture a proxy for skill importance by measuring the comparative advantage of each skill in each job ad, as seen in the $RCA$ equation in \textit{Materials \& Methods}.
Our measure controls for high-occurring skills through normalization and develops a measure of skill importance within individual job ads that later represent skill importance within labor market groups (occupations, industries etc.).


\begin{thebibliography}{10}

\bibitem{acemoglu2011skills}
Acemoglu D, Autor D.
\newblock Skills, tasks and technologies: Implications for employment and
  earnings.
\newblock In: Handbook of Labor Economics. vol.~4. Elsevier; 2011. p.
  1043--1171.

\bibitem{brynjolfsson2014second}
Brynjolfsson E, McAfee A.
\newblock The second machine age: Work, progress, and prosperity in a time of
  brilliant technologies.
\newblock WW Norton \& Company; 2014.

\bibitem{schwab2017fourth}
Schwab K.
\newblock The Fourth Industrial Revolution.
\newblock Currency; 2017.

\bibitem{Frey2017}
Frey CB, Osborne MA.
\newblock The Future of Employment: How susceptible are jobs to
  computerisation?
\newblock Technological Forecasting and Social Change. 2017;114:254--280.

\bibitem{acemoglu2018artificial}
Acemoglu D, Restrepo P.
\newblock Artificial Intelligence, Automation and Work.
\newblock National Bureau of Economic Research; 2018.

\bibitem{Frank2019-dn}
Frank MR, Autor D, Bessen JE, Brynjolfsson E, Cebrian M, Deming DJ, et~al.
\newblock Toward understanding the impact of artificial intelligence on labor.
\newblock Proceedings of the National Academy of Sciences of the USA. 2019; p.
  201900949.

\bibitem{sila2019job}
Sila U.
\newblock {Job displacement in Australia: Evidence from the HILDA survey}.
\newblock OECD; 2019.

\bibitem{Abs2020-ws}
{ABS}. Labour {F}orce, {A}ustralia; 2020.
\newblock
  \url{https://www.abs.gov.au/statistics/labour/employment-and-unemployment/labour-force-australia/latest-release}.

\bibitem{nedelkoska2015skill}
Nedelkoska L, Neffke F, Wiederhold S.
\newblock Skill mismatch and the costs of job displacement.
\newblock In: Annual Meeting of the American Economic Association; 2015.

\bibitem{poletaev2008human}
Poletaev M, Robinson C.
\newblock Human capital specificity: evidence from the Dictionary of
  Occupational Titles and Displaced Worker Surveys, 1984--2000.
\newblock Journal of Labor Economics. 2008;26(3):387--420.

\bibitem{gathmann2010general}
Gathmann C, Sch{\"o}nberg U.
\newblock How general is human capital? {A} task-based approach.
\newblock Journal of Labor Economics. 2010;28(1):1--49.

\bibitem{nedelkoska2018}
Nedelkoska L, Diodato D, Neffke F.
\newblock Is our human capital general enough to withstand the current wave of
  technological change?
\newblock Center for International Development, Harvard University; 2018.

\bibitem{MovingBetweenJobs}
Bechichii N, Grundkei R, Jameti S, Squicciarini M.
\newblock Moving Between Jobs: An Analysis of Occupation Distances and Skill
  Needs.
\newblock OECD; 2018. 52.

\bibitem{HILDA_2020}
{Department of Social Services and Melbourne Institute of Applied Economic and
  Social Research}. The Household, Income and Labour Dynamics in Australia
  ({HILDA}) Survey, {RESTRICTED} {RELEASE} 18 (Waves 1-18); 2020.

\bibitem{Wef2018-eu}
{WEF}.
\newblock Towards a Reskilling Revolution A Future of Jobs for All.
\newblock World Economic Forum and The Boston Consulting Group; 2018.

\bibitem{Wef2019-jv}
{WEF}.
\newblock Towards a Reskilling Revolution {Industry-Led} Action for the Future
  of Work.
\newblock World Economic Forum and The Boston Consulting Group; 2019.

\bibitem{Reskilling-AUS-2019}
{Australia  Department of Employment, Skills, Small and Family Business  Future
  of Work Taskforce }.
\newblock Reskilling {A}ustralia: A data-driven approach.
\newblock Australian Government; 2019.

\bibitem{Kyle_Demaria2020-ec}
Kyle~Demaria KF, Wardrip K.
\newblock Exploring a Skills-based Approach to Occupational Mobility.
\newblock Federal Reserve Banks of Philadelphia, and Cleveland; 2020.

\bibitem{abc-lockdown}
ABC.
\newblock Australia's social distancing rules have been enhanced to slow
  coronavirus --- here's how they work.
\newblock ABC News. 2020;.

\bibitem{US_Department_of_Labor_undated-oe}
{U S  Department of Labor}. {O*NET}; 2020.
\newblock \url{https://www.onetonline.org/}.

\bibitem{Australian_Bureau_of_Statistics2019-sv}
{Australian Bureau of Statistics}. 6291.0.55.003 - {Labour Force, Australia,
  Detailed, Quarterly}; 2019.

\bibitem{Hidalgo2007-qk}
Hidalgo CA, Klinger B, Barab{\'a}si AL, Hausmann R.
\newblock The product space conditions the development of nations.
\newblock Science. 2007;317(5837):482--487.

\bibitem{Vollrath1991-kr}
Vollrath TL.
\newblock A theoretical evaluation of alternative trade intensity measures of
  revealed comparative advantage.
\newblock Weltwirtsch Arch. 1991;127(2):265--280.

\bibitem{Shutters2016-fe}
Shutters ST, Muneepeerakul R, Lobo J.
\newblock Constrained pathways to a creative urban economy.
\newblock Urban Studies. 2016;53(16):3439--3454.

\bibitem{Alabdulkareem2018-jl}
Alabdulkareem A, Frank MR, Sun L, AlShebli B, Hidalgo C, Rahwan I.
\newblock Unpacking the polarization of workplace skills.
\newblock Science Advances. 2018;4(7):eaao6030.

\bibitem{dawson2019adaptively}
Dawson N, Rizoiu MA, Johnston B, Williams MA.
\newblock Adaptively selecting occupations to detect skill shortages from
  online job ads.
\newblock In: 2019 IEEE International Conference on Big Data (Big Data). IEEE;
  2019. p. 1637--1643.

\bibitem{XGBoost}
Chen T, Guestrin C.
\newblock {XGBoost}: A Scalable Tree Boosting System.
\newblock In: Proceedings of the 22nd {ACM} {SIGKDD} International Conference
  on Knowledge Discovery and Data Mining. KDD '16. New York, NY, USA:
  Association for Computing Machinery; 2016. p. 785--794.

\bibitem{robinson2018occupational}
Robinson C.
\newblock Occupational mobility, occupation distance, and specific human
  capital.
\newblock Journal of Human Resources. 2018;53(2):513--551.

\bibitem{bergstra2012random}
Bergstra J, Bengio Y.
\newblock Random search for hyper-parameter optimization.
\newblock {Journal of Machine Learning Research}. 2012;13(Feb):281--305.

\bibitem{2019TRIMAP}
{Amid} E, {Warmuth} MK.
\newblock {TriMap: Large-scale Dimensionality Reduction Using Triplets}.
\newblock ArXiv e-prints. 2019;.

\bibitem{brynjolfsson2013complementarity}
Brynjolfsson E, Milgrom P.
\newblock Complementarity in organizations.
\newblock {The Handbook of Organizational Economics}. 2013; p. 11--55.

\bibitem{autor2013update}
Autor D, Price B.
\newblock The Changing Task Composition of the {US} Labor Market: {An} update
  of {Autor}, {Levy}, and {Murnane} (2003).
\newblock MIT Paper. 2013;21.

\bibitem{Harris2020-dt}
{Harris, Rob and Bagshaw, Eryk}.
\newblock Strict new controls announced as Morrison government tries to limit
  spread of {COVID-19}.
\newblock {The Sydney Morning Herald}. 2020;.

\bibitem{Faethm_data}
{Faethm}. Australian essential and non-essential occupations during {COVID-19}
  2020; 2021.
\newblock
  \url{https://github.com/Faethm-ai/open-data/blob/main/essential-occupations-AUS/ANZSCO_4digit_essential_v_nonessential.csv}.

\bibitem{bresnahan2002information}
Bresnahan TF, Brynjolfsson E, Hitt LM.
\newblock Information technology, workplace organization, and the demand for
  skilled labor: Firm-level evidence.
\newblock The Quarterly Journal of Economics. 2002;117(1):339--376.

\bibitem{Bessen2015-pp}
Bessen J.
\newblock Learning by Doing: The Real Connection between Innovation, Wages, and
  Wealth.
\newblock Yale University Press; 2015.

\bibitem{Amazon2019}
Koehn E.
\newblock `We're just getting started': Amazon {A}ustralia revenue surges to
  \$292m.
\newblock The Sydney Morning Herald. 2019;.

\bibitem{watson2004sample}
Watson N, Wooden M.
\newblock {Sample attrition in the HILDA survey}.
\newblock {Australian Journal of Labour Economics}. 2004;7(2):293--308.

\bibitem{moro2021universal}
Moro E, Frank MR, Pentland A, Rutherford A, Cebrian M, Rahwan I.
\newblock Universal resilience patterns in labor markets.
\newblock {Nature Communications}. 2021;12(1):1--8.

\bibitem{Kern2019-zg}
Kern ML, McCarthy PX, Chakrabarty D, Rizoiu MA.
\newblock Social media-predicted personality traits and values can help match
  people to their ideal jobs.
\newblock Proceedings of the National Academy of Sciences of the United States
  of America. 2019;116(52):26459--26464.

\bibitem{browne2003intersection}
{Browne, Irene and Misra, Joya}.
\newblock The intersection of gender and race in the labor market.
\newblock {Annual Review of Sociology}. 2003;29(1):487--513.

\bibitem{pager2009discrimination}
{Pager, Devah and Bonikowski, Bart and Western, Bruce}.
\newblock Discrimination in a low-wage labor market: A field experiment.
\newblock {American Sociological Review}. 2009;74(5):777--799.

\bibitem{carlsson2019age}
{Carlsson, Magnus and Eriksson, Stefan}.
\newblock {Age discrimination in hiring decisions: Evidence from a field
  experiment in the labor market}.
\newblock {Labour Economics}. 2019;59:173--183.

\bibitem{foley2018does}
{Foley, Meraiah and Williamson, Sue}.
\newblock Does anonymising job applications reduce gender bias? Understanding
  managers’ perspectives.
\newblock {Gender in Management: An International Journal}. 2018;.

\bibitem{OECD2020}
OECD.
\newblock OECD Employment Outlook 2020; 2020.
\newblock Available from:
  \url{https://www.oecd-ilibrary.org/content/publication/1686c758-en}.

\bibitem{borjas2010labor}
Borjas GJ, Van~Ours JC.
\newblock Labor economics.
\newblock McGraw-Hill/Irwin Boston; 2010.

\bibitem{Schultz1961-uq}
Schultz TW.
\newblock Investment in Human Capital.
\newblock The American Economic Review. 1961;51(1):1--17.

\bibitem{becker1964human}
Becker G.
\newblock Human capital.
\newblock Columbia University: Columbia University Press; 1964.

\bibitem{Becker1990-zc}
Becker GS, Murphy KM, Tamura R.
\newblock Human Capital, Fertility, and Economic Growth.
\newblock Journal of Political Economy. 1990;98(5, Part 2):S12--S37.

\bibitem{pries2005}
Pries M, Rogerson R.
\newblock Hiring policies, labor market institutions, and labor market flows.
\newblock Journal of Political Economy. 2005;113(4):811--839.

\bibitem{bassanini2013}
Bassanini A, Garnero A.
\newblock Dismissal protection and worker flows in OECD countries: Evidence
  from cross-country/cross-industry data.
\newblock Labour Economics. 2013;21:25--41.

\bibitem{hassler2005}
Hassler J, Rodriguez~Mora JV, Storesletten K, Zilibotti F.
\newblock A positive theory of geographic mobility and social insurance.
\newblock International Economic Review. 2005;46(1):263--303.

\bibitem{goldin2016}
Goldin CD.
\newblock In: Human Capital. Heidelberg, Germany: Springer Verlag; 2016.

\bibitem{nedelkoska2019}
Nedelkoska L, Neffke F.
\newblock Skill Mismatch and Skill Transferability: Review of Concepts and
  Measurements.
\newblock Papers in Evolutionary Economic Geography. 2019;.

\bibitem{wasmer2006}
Wasmer E.
\newblock General versus specific skills in labor markets with search frictions
  and firing costs.
\newblock American Economic Review. 2006;96(3):811--831.

\bibitem{Oecd2019-cl}
{OECD}.
\newblock {OECD} Skills Strategy 2019 - Skills to Shape a Better Future.
\newblock OECD; 2019.

\bibitem{Gardiner2018-dt}
Gardiner A, Aasheim C, Rutner P, Williams S.
\newblock Skill Requirements in Big Data: A Content Analysis of Job
  Advertisements.
\newblock Journal of Computer Information Systems. 2018;58(4):374--384.

\bibitem{topel1992}
Topel RH, Ward MP.
\newblock Job mobility and the careers of young men.
\newblock The Quarterly Journal of Economics. 1992;107(2):439--479.

\bibitem{freeman1975}
Freeman RB.
\newblock Overinvestment in college training?
\newblock Journal of human resources. 1975; p. 287--311.

\bibitem{goldin2009}
Goldin CD, Katz LF.
\newblock The race between education and technology.
\newblock Harvard University Press; 2009.

\bibitem{vona2015innovation}
Vona F, Consoli D.
\newblock Innovation and skill dynamics: a life-cycle approach.
\newblock Industrial and Corporate Change. 2015;24(6):1393--1415.

\bibitem{mincer1991human}
Mincer J.
\newblock Human capital, technology, and the wage structure: what do time
  series show?
\newblock National Bureau of Economic Research; 1991.

\bibitem{berman1994changes}
Berman E, Bound J, Griliches Z.
\newblock Changes in the demand for skilled labor within US manufacturing:
  evidence from the annual survey of manufactures.
\newblock The Quarterly Journal of Economics. 1994;109(2):367--397.

\bibitem{autor1998computing}
Autor DH, Katz LF, Krueger AB.
\newblock Computing inequality: have computers changed the labor market?
\newblock The Quarterly journal of economics. 1998;113(4):1169--1213.

\bibitem{autor2013putting}
Autor DH, Handel MJ.
\newblock Putting tasks to the test: Human capital, job tasks, and wages.
\newblock Journal of labor Economics. 2013;31(S1):S59--S96.

\bibitem{goos2014explaining}
Goos M, Manning A, Salomons A.
\newblock Explaining job polarization: Routine-biased technological change and
  offshoring.
\newblock American economic review. 2014;104(8):2509--26.

\bibitem{Brown2020-yu}
Brown TB, Mann B, Ryder N, Subbiah M, Kaplan J, Dhariwal P, et~al.
\newblock Language Models are {Few-Shot} Learners.
\newblock In: Advances in Neural Information Processing Systems (NeurIPS 2020);
  2020.

\bibitem{Touvron2020-mz}
Touvron H, Vedaldi A, Douze M, J{\'e}gou H.
\newblock Fixing the train-test resolution discrepancy: {FixEfficientNet};
  2020.

\bibitem{Silver2018-kt}
Silver D, Hubert T, Schrittwieser J, Antonoglou I, Lai M, Guez A, et~al.
\newblock A general reinforcement learning algorithm that masters chess, shogi,
  and Go through self-play.
\newblock Science. 2018;362(6419):1140--1144.

\bibitem{blinder2013alternative}
Blinder AS, Krueger AB.
\newblock Alternative measures of offshorability: a survey approach.
\newblock Journal of Labor Economics. 2013;31(S1):S97--S128.

\bibitem{Autor2013-we}
Autor DH, Dorn D, Hanson GH.
\newblock The China Syndrome: Local Labor Market Effects of Import Competition
  in the United States.
\newblock Am Econ Rev. 2013;103(6):2121--2168.

\bibitem{shaw1984formulation}
Shaw KL.
\newblock A formulation of the earnings function using the concept of
  occupational investment.
\newblock Journal of Human Resources. 1984; p. 319--340.

\bibitem{shaw1987occupational}
Shaw KL.
\newblock Occupational change, employer change, and the transferability of
  skills.
\newblock Southern Economic Journal. 1987; p. 702--719.

\bibitem{ingram2006returns}
Ingram BF, Neumann GR.
\newblock The returns to skill.
\newblock Labour economics. 2006;13(1):35--59.

\bibitem{bechichi2018}
Bechichi N, Grundke R, Jamet S, Squicciarini M.
\newblock Moving between jobs; 2018.

\bibitem{grundke2017having}
Grundke R, Jamet S, Kalamova M, Squicciarini M.
\newblock Having the right mix: The role of skill bundles for comparative
  advantage and industry performance in GVCs; 2017.

\bibitem{Bessen2002}
Bessen J.
\newblock Technology adoption costs and productivity growth: The transition to
  information technology.
\newblock Review of Economic Dynamics. 2002;.

\bibitem{Bessen2018}
Bessen JE, Impink SM, Seamans R, Reichensperger L.
\newblock The Business of {AI} Startups; 2018.

\bibitem{rogers1976new}
Rogers EM.
\newblock New product adoption and diffusion.
\newblock Journal of consumer Research. 1976;2(4):290--301.

\bibitem{karahanna1999information}
Karahanna E, Straub DW, Chervany NL.
\newblock Information technology adoption across time: a cross-sectional
  comparison of pre-adoption and post-adoption beliefs.
\newblock MIS quarterly. 1999; p. 183--213.

\bibitem{im2011international}
Im I, Hong S, Kang MS.
\newblock An international comparison of technology adoption: Testing the UTAUT
  model.
\newblock Information \& management. 2011;48(1):1--8.

\bibitem{thong1999integrated}
Thong JY.
\newblock An integrated model of information systems adoption in small
  businesses.
\newblock Journal of management information systems. 1999;15(4):187--214.

\bibitem{andres2010diffusion}
Andr{\'e}s L, Cuberes D, Diouf M, Serebrisky T.
\newblock The diffusion of the Internet: A cross-country analysis.
\newblock Telecommunications policy. 2010;34(5-6):323--340.

\bibitem{perrin2015social}
Perrin A.
\newblock Social media usage.
\newblock Pew research center. 2015; p. 52--68.

\bibitem{Bughin2018-qr}
Bughin J, Seong J, Manyika J, Chui M, Joshi R.
\newblock Notes from the {AI} frontier: Modeling the impact of {AI} on the
  world economy.
\newblock McKinsey Global Institute; 2018.

\bibitem{moorthy1985using}
Moorthy KS.
\newblock Using game theory to model competition.
\newblock Journal of Marketing Research. 1985;22(3):262--282.

\bibitem{Andrews2015-yy}
Andrews D, Criscuolo C, Gal PN.
\newblock Frontier firms, technology diffusion and public policy: Micro
  evidence from {OECD} countries.
\newblock OECD; 2015.

\bibitem{Mamer1987-cn}
Mamer JW, McCardle KF.
\newblock Uncertainty, Competition, and the Adoption of New Technology.
\newblock Management Science. 1987;33(2):161--177.

\bibitem{hall2003adoption}
Hall BH, Khan B.
\newblock Adoption of new technology.
\newblock National bureau of economic research; 2003.

\bibitem{ABStech}
Business Use of Information Technology; 2017.
\newblock
  \url{https://www.abs.gov.au/statistics/industry/technology-and-innovation/business-use-information-technology/latest-release}.

\bibitem{beaudry2006endogenous}
Beaudry P, Doms M, Lewis E.
\newblock Endogenous skill bias in technology adoption: City-level evidence
  from the IT revolution.
\newblock National Bureau of Economic Research; 2006.

\bibitem{Andrews2018-wt}
Andrews D, Nicoletti G, Timiliotis C.
\newblock Digital technology diffusion: A matter of capabilities, incentives or
  both?
\newblock OECD; 2018.

\bibitem{anderson2004information}
Anderson ST, Newell RG.
\newblock Information programs for technology adoption: the case of
  energy-efficiency audits.
\newblock Resource and Energy economics. 2004;26(1):27--50.

\bibitem{Brynjolfsson2018-rs}
Brynjolfsson E, Rock D, Syverson C.
\newblock Artificial Intelligence and the Modern Productivity Paradox: A Clash
  of Expectations and Statistics.
\newblock In: The Economics of Artificial Intelligence: An Agenda. University
  of Chicago Press; 2018.

\bibitem{perino2012does}
Perino G, Requate T.
\newblock Does more stringent environmental regulation induce or reduce
  technology adoption? When the rate of technology adoption is inverted
  U-shaped.
\newblock Journal of Environmental Economics and Management.
  2012;64(3):456--467.

\bibitem{GoogleCloud}
Cloud {AutoML};.
\newblock \url{https://cloud.google.com/automl}.

\bibitem{ANZSCO2013}
{Australian Bureau of Statistics}. 1220.0 - {ANZSCO} -- {Australian and New
  Zealand Standard Classification of Occupations, 2013, Version 1.2}; 2013.
\newblock
  \url{https://www.abs.gov.au/ausstats/abs@.nsf/0/E3031B89999B4582CA2575DF002DA702?opendocument#:~:text=The\%20structure\%20of\%20ANZSCO\%20has,grouped\%20into\%20'minor\%20groups'.}

\bibitem{ONET_ANZSCO_concordance}
{Australian Federal Department of Education, Skills and Employment}. {ANZSCO}
  to {O*NET} concordance;.

\bibitem{Carnevale2014-xc}
Carnevale A, Jayasundera T, Repnikov D.
\newblock Understanding Online Job Ads Data.
\newblock Georgetown University; 2014.

\end{thebibliography}
\end{document}